\documentclass[preprint]{aastex}

\usepackage{graphicx}
\newcommand\gtsim{\mathrel{\lower0.6ex\hbox{$\buildrel {\textstyle >}
   \over {\scriptstyle \sim}$}}}
\newcommand\ltsim{\mathrel{\lower0.6ex\hbox{$\buildrel {\textstyle <}
   \over {\scriptstyle \sim}$}}}
\usepackage{aas_macros}
\usepackage[footnotesize]{subfigure}

\slugcomment{To appear in Ap.J.}
\shorttitle{3C~293}
\shortauthors{Floyd et al.}

\begin{document}

\title{The jet and circumnuclear environment of 3C~293
\footnote{Based on observations with the NASA/ESA Hubble Space Telescope, obtained at the Space Telescope Science Institute, which is operated by the Association of Universities for Research in Astronomy, Inc. (AURA), under NASA contract NAS5-26555}
\footnote{Based on observations with MERLIN, a national facility operated by the University of Manchester on behalf of PPARC in the UK.}}

\author{David J. E. Floyd}
\affil{Space Telescope Science Institute, 3700 San Martin Drive, Baltimore, MD 21218, U.S.A.}
\email{floyd@stsci.edu}

\author{Eric Perlman}
\affil{Joint Center for Astrophysics, Physics Dept. University of Maryland, Baltimore County, 1000 Hilltop Circle, Baltimore, MD 21250}

\author{J. Patrick Leahy, R. J. Beswick, Neal J. Jackson}
\affil{University of Manchester, Jodrell Bank Observatory, Macclesfield, Cheshire, England} 

\author{William B. Sparks}
\affil{Space Telescope Science Institute, 3700 San Martin Drive, Baltimore, MD 21218, U.S.A.}

\author{David J. Axon, Christopher P. O'Dea}
\affil{Department of Physics, Rochester Institute of Technology, 54 Lomb Memorial Drive, Rochester, NY 14623-5604}

\begin{abstract}
We present the new HST near-infrared polarimetry, broad and narrow-band imaging, and MERLIN 4.5GHz Multi-Frequency Synthesis radio imaging of 3C~293, a unique radio galaxy whose host is an obvious merger remnant, in an exceptionally under-dense region of space. We have discovered near-infrared, optical, and ultra-violet synchrotron emission from the jet. In the optical, the jet is mostly obscured by a dust lane, but three knots are clear in our HST NICMOS images at 1.6 and 2.0~$\mu$m, clearly aligning with features in the radio. The outer jet knot is highly polarized ($\approx 15\%$) at 2~$\mu$m, confirming the synchrotron emission mechanism. The radio-IR spectral index steepens significantly with distance from the nucleus, as in 3C~273 and in contrast to M~87. The inner knot is visible (with hindsight) on the WFPC2 and STIS images obtained for the earlier 3CR HST snapshot surveys. There is no [Fe~{\sc ii}] emission seen associated with the jet, constraining the role of shock-induced ionisation by the jet. Overall there is a strong implication that the NIR jet emission is indeed synchrotron.

From our NIR images, the core of the galaxy is clearly identifiable with the main feature in the western extension of the radio ``jet'' image, although no unresolved AGN component is identifiable even at $K$-band, consistent with an FRII-like nucleus obscured by an optically thick torus. The galaxy appears to have a single nucleus, with any multiple nuclei falling within the central $\ltsim100$~pc. 

\end{abstract}


\newpage

\section{Introduction}
\label{sec-intro}
One of the greatest gaps in our understanding of active galaxies (and arguably of galaxy evolution in general) is the trigger for the onset of the active phase. It is not yet obvious why at any given time, a small fraction of galaxies have a degree of central activity orders of magnitude larger than that expected of normal galaxies, nor why a still smaller fraction produce radio jets.  The majority of radio galaxies (RGs) are clearly ellipticals, although they often show morphological peculiarities, especially kpc-scale disks (\citealt{heckman+85a,martel+99}, Floyd et al. in preparation). Much work has been done investigating the possibility that active galaxies form as a result of mergers~\citep{toomretoomre72,hutchings+84,wilsoncolbert95,canalizostockton01}. The evidence for this includes disturbed morphologies such as tidal tails. More recently, {\em Hubble Space Telescope} (HST) imaging of a large sample of 3CR galaxies~\citep{dekoff+00} has revealed that a large number of nearby radio-loud elliptical galaxies contain prominent dust lanes, suggestive of the capture of interstellar medium by the accreting galaxy. Furthermore,~\citet{capetti+02} have shown that radio loudness correlates with the inner structure of elliptical galaxies, implying a direct link with mergers. 
~\citet{perlman+02} found kpc-scale disks present in many compact steep spectrum objects (CSO's), which were hosted by otherwise normal ellipticals. However, in this case, it was found that the isophotal anomalies were too small to suggest interactions within the last $<10^{8}$~yr.
It remains unclear whether such anomalies distinguish radio galaxies from the elliptical galaxy population in general, or are simply a legacy of that population, as statistical comparisons of radio galaxy and quiescent elliptical galaxy morphology at high spatial resolution are limited.

\begin{figure}[ht]
\epsscale{0.8}\plotone{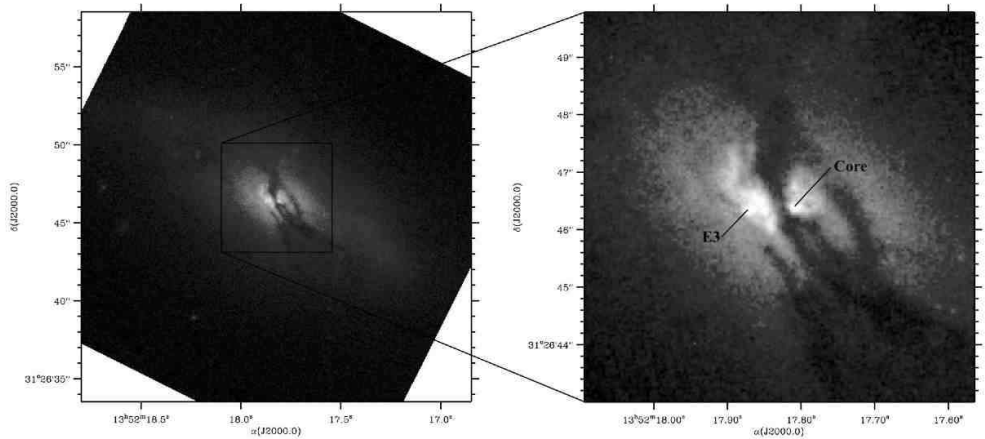}
\caption{\label{fig-optical} WFPC2 {\sc f702w} ($R$-band) image of 3C~293~\citep{martel+99}. The innermost knot E3 (following the labelling of Beswick et al. 2004) and the core are marked, with 5GHz MERLIN radio contours overlayed for reference.}
\end{figure}

\begin{figure}[ht]
\epsscale{0.4}\plotone{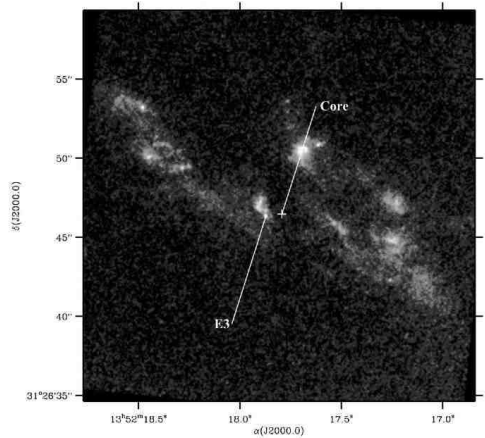}
\caption{\label{fig-uv} STIS NUV ({\sc f25srf2}) image of 3C~293~\citep{allen+02}. The innermost knot E3 (following the labelling of Beswick et al. 2004), and the position of the core are indicated.}
\end{figure}

In contrast to the majority of RG's the host galaxy of 3C~293 (UGC~8782) has variously been classified as a Spiral (in the UGC), Sb~\citep{sandage66}, D5-6~\citep{burbidgecrowne79}, S0~\citep{colla+75}, and an Elliptical (\citealt{VCV01} and references therein). 
This confusion is readily understandable in the light of our recent HST/WFPC2 snapshot image (\citealt{martel+99} - reproduced here in Fig.~\ref{fig-optical}), which shows that it has the most disturbed optical morphology of any low-redshift 3CR source. Presumably not by coincidence, 3C~293 is one of the best examples of interactions between jets from the AGN and cool ($\ll10^{6}$K) gas in the interstellar medium. The disturbed morphology and interesting radio properties of this object have made for a challenging and rewarding target for numerous studies at multiple wavelengths over the last two decades.

The radio source in 3C~293 contains structure on two distinct scales, strongly suggesting a sequence of discrete outbursts. On the largest (over 200~kpc) scales, the object resembles a moderately large FRII (see observations by ~\citealt{bridle+81,vanbreugel+84}) except that it has an unusually large amount of flux from an extended, steep-spectrum core that has been resolved into a jet structure~\citep{bridle+81,akujor+96,beswick+02,beswick+04}. The radio jets run almost east-west  from the nucleus, but twist by $\sim30^{\circ}$ at $10\arcsec$ distance from the nucleus. The core emission is steeply inverted ($\alpha^{15}_{22}=-1.05$~\citealt{akujor+96}; $\alpha^{1.3}_{4.5}=0.11$~\citealt{beswick+04}, following the convention that $F_{\nu}\propto\nu^{-\alpha}$), and is still detectable at 2.7mm. 

UGC~8782 is listed as a possible, but unlikely, outlying member of the $z<0.05$ cluster 1352.0+3107~\citep{burnsowen77}, being 50\arcmin~from the center of the cluster. Moreover,~\citet{stocke79} notes it to be in a region of ``exceptionally low galaxy number density.'' 
The galaxy appears to be interacting with a small companion some $37\arcsec$ ($\sim30$~kpc) to the southwest, which unfortunately falls in the ``missing segment'' of the WFPC2 field, and off the edge of the small NIC2 chip. The companion is joined to the main galaxy by a faint ($\mu_{V}\approx23-23.5$~mag~arcsec$^{-2}$) bridge of optical emission (clearly seen extending to the SW on our images), with a still fainter ($\mu_{V}\approx24.5$~mag~arcsec$^{-2}$) fan or tail of optical emission extending $50-60\arcsec$ beyond the companion~\citep{heckman+86}. Both the companion and tidal tail are visible on the 48-inch Palomar Schmidt DSS plates (see for example~\citealt{evans+99}). \citet{emonts+05a} find a total of 3 sources in H~{\sc i} emission, within a much larger radius of $\sim100$~kpc, in both angular separation and velocity.
UGC~8782 is bluer than normal ellipticals, suggesting a population of young stars~\citep{smithheckman89}. Furthermore the UV morphology (Fig.~\ref{fig-uv} and~\citealt{allen+02}) is indicative of dust-associated star-formation, with diffuse filamentary emission along the same P.A. as the optical dust disk. 

Dynamically, UGC~8782 is found to be an oblate rotator, rather than a genuine spiral as it does not have a low velocity-dispersion stellar disk~\citep{heckman+85b}. It does however have a well-defined gaseous disk, clearly detected in the optical HST image (Fig.~\ref{fig-optical}) as a prominent complex of dust lanes that cross the major axis of the galaxy at an oblique angle. This dust lane has also been observed through neutral hydrogen absorption~\citep{baanhaschick81,haschickbaan85,beswick+02,beswick+04}, optical emission lines~\citep{vanbreugel+84}, and in CO absorption~\citep{evans+99}. \citet{morganti+03} report an additional very broad blue-shifted absorption component against the central regions, suggesting an unusually strong outflow of neutral gas from the AGN, which has more recently been localised to the eastern side of the source by~\citet{emonts+05b}. These three absorption components are discussed in greater detail in section~\ref{sec-dust}

Furthermore, 3C~293 is one of the few FRII radio sources to have been detected in CO emission~\citep{evans+99}. The CO $J=1\rightarrow0$ (115.271~GHz) emission is extended over an asymmetric region $7\arcsec$ across, and roughly centered on the unresolved continuum source. The high concentration ($\sim10^{10}$~M$_{\odot}$, an order of magnitude more molecular gas than found in other comparable FRII's, e.g. PKS 0634-206~\citealt{odea+94}) of molecular gas in the central few kpc (Fig.~ 4 of Evans et al. 1999) supports the idea that 3C~293 has recently been involved in a gas-rich galaxy-galaxy interaction or merger, in which large volumes of gas and material have been cannibalised and are being used as fuel for the AGN. The CO component is moderately broad ($\Delta v_{\mathrm FWHM} \sim 400$~km~s$^{-1}$), with a narrow ($\Delta v_{\mathrm abs} \sim 60$~km~s$^{-1}$) absorption feature (consistent with the narrow H~{\sc i} absorption) superimposed along the line-of-sight to the continuum source. The structure and kinematics of the CO emission indicate a disk of molecular gas that is rotating around the continuum source (moving away from us on the Eastern side). The continuum source itself is consistent with a power-law extrapolation of the radio core flux densities, and is interpreted by Evans et al. as the AGN and circum-nuclear accretion disk, and not from dust associated with the molecular gas. 

Most of these features (CO emission aside) are shared with Centaurus A and also with 3C~305, and the favored interpretation for all three is that these are recent mergers between a large elliptical and a small spiral galaxy (e.g.~\citealt{eb83}). In Cen A, the radio jets escape approximately perpendicular to the gas disk, resulting in minimal interaction; in contrast, in the other two, the jets and disk axes are severely misaligned and the jets appear to directly impact on the cold gas left behind by the merger.

We assume throughout $H_{0}=70$~km~s$^{-1}$Mpc$^{-1}$. The optical counterpart of 3C~293 (UGC~8782) is at a redshift of $z=0.04503$, placing it at a distance of $D=180$Mpc. At this distance, $1\arcsec$ corresponds to a linear scale of 0.815~kpc.

\section{Observations and data reduction}
\subsection{HST}
\subsubsection{NICMOS observations}
We observed 3C~293 with the NIC2 camera aboard HST, using the {\sc f160w} filter (1.61~$\mu$m), the long-wavelength polarizers (2.00~$\mu$m), and the medium-band {\sc f171m} (1.72~$\mu$m; $\Delta\lambda=248$~\AA) and {\sc f180m} (1.80~$\mu$m; $\Delta\lambda=239$~\AA) filters on 1998 August 19 (see table~\ref{tab-obs}). 

We used a two-pointing dither and {\sc multiaccum} mode in order to eliminate most of the detector artefacts. Data were processed using the NICMOS pipeline up to the {\sc calnica} procedure, which produces a calibrated image from each {\sc multiaccum} pointing. At this point the image data quality mask was updated with a near-contemporary mask for the various bad pixels in NIC2. We also masked the coronographic hole and the ``photometrically challenged column.''

\begin{figure*}[ht]
\epsscale{0.8}\plotone{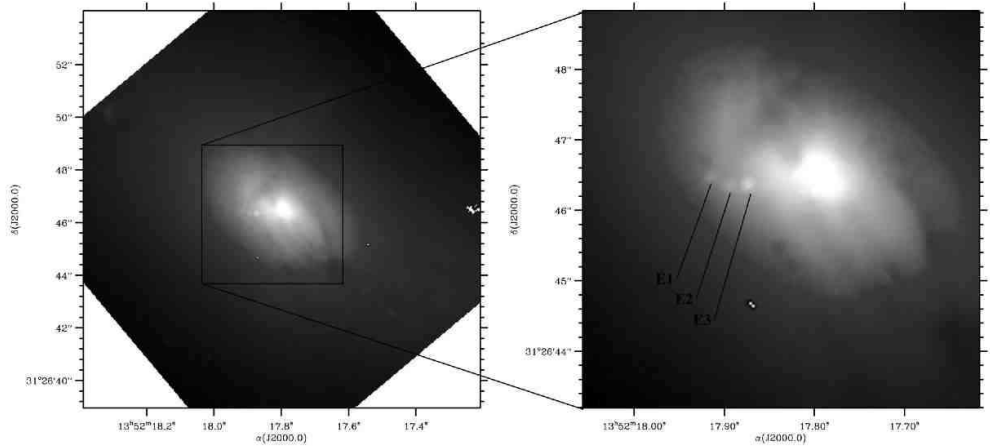}
\caption{\label{fig-H} The {\sc f160w} (1.6~$\mu$m) image of 3C~293. The inset shows the region of the nucleus and the infrared jet counterpart, with the three distinct knots labelled following the convention of Beswick et al. (2004). }
\end{figure*}

\begin{figure*}[ht!]
\epsscale{0.8}\plotone{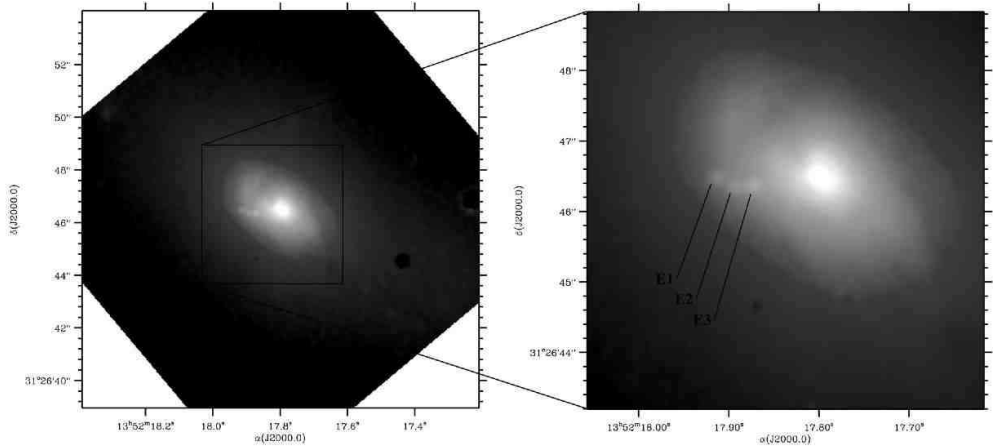}
\caption{\label{fig-K} The $\lambda2.0~\mu$m total intensity image made from the long wavelength polarizer images. The inset shows the region of the nucleus and the infrared jet counterpart, with the three distinct knots labelled following the convention of Beswick et al. (2004). }
\end{figure*}

NICMOS images are affected by a ``pedestal'' or unpredictable bias, which may be different in the four quadrants of the image. We used {\sc pedsub} in {\sc iraf (stsdas.hst\_calib.nicmos)} to remove this effect. Results were checked for each quadrant, and the overall sky background subtracted, using the {\sc msky2} program of Mark Dickinson. 
Dithering was performed using the standard {\sc drizzle} package~\citep{fruchterhook02} in {\sc iraf}. We used a scale factor of 0.5 to give a final pixel scale of $0.038\arcsec$ in our final mosaiced images (Fig.~\ref{fig-H},~\ref{fig-K}).


\subsubsection{Optical-UV archival data}
In addition we obtained WFPC2/PC ({\sc f702w}) and STIS/ NUV MAMA (Strontium Fluoride filter {\sc f25srf2}) images from our snapshot surveys of the 3CR, originally published in~\citet{martel+99,allen+02} respectively. See those papers for the appropriate data reduction techniques.

\subsubsection{Infrared Polarimetry}
\begin{figure*}[htf]
\epsscale{0.7}\plotone{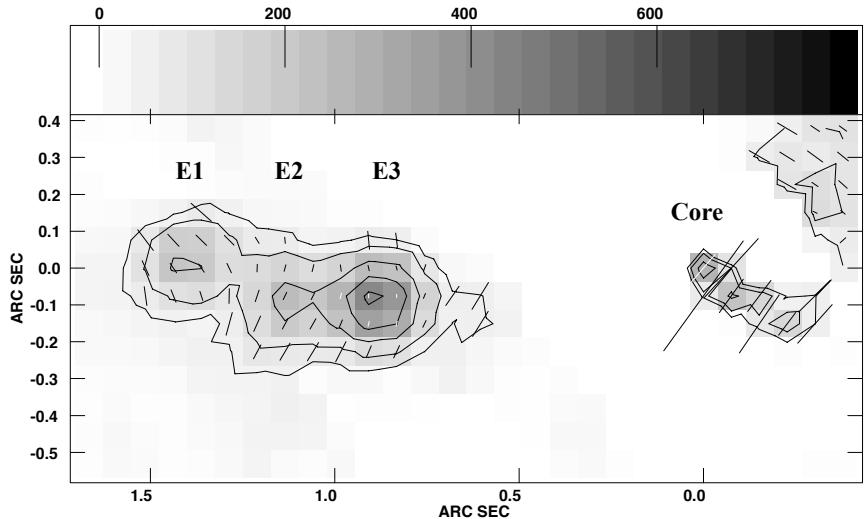}
\caption{\label{fig-IRpol} $2~\mu$m polarization map (B-vectors) overlayed with the $2~\mu$m galaxy model-subtracted contours and grey-scale (in milli-counts per second). See section~\ref{sec-hostmodel} for details of the galaxy modelling. The contours are shown at flux levels of $(50, 100, 200, 300, 400)\times10^{-03}$ counts~s$^{-1}$. A $0\arcsec.1$ polarization vector corresponds to a polarization fraction of 10\%. The jet is aligned roughly east-west, but note the slight northward bend. The strong ($\sim30^\circ$) twist in the radio axis occurs much further out at $\approx10$\arcsec from the nucleus.}
\end{figure*}

The images taken with the three polarizers were combined to form Stokes $I, Q$, and $U$ parameters, using the {\sc polarize} IDL software~\citep{hines+00}, available on the NICMOS website. The Stokes $I$ image is presented in Fig.~\ref{fig-K}. The $Q$ and $U$ parameters were used to create polarization maps in {\sc aips}, with contours overlayed from the model-subtracted unpolarized image (see section~\ref{sec-hostmodel} and Fig.~\ref{fig-IRpol}).

\begin{figure*}[htf]
\epsscale{0.7}\plottwo{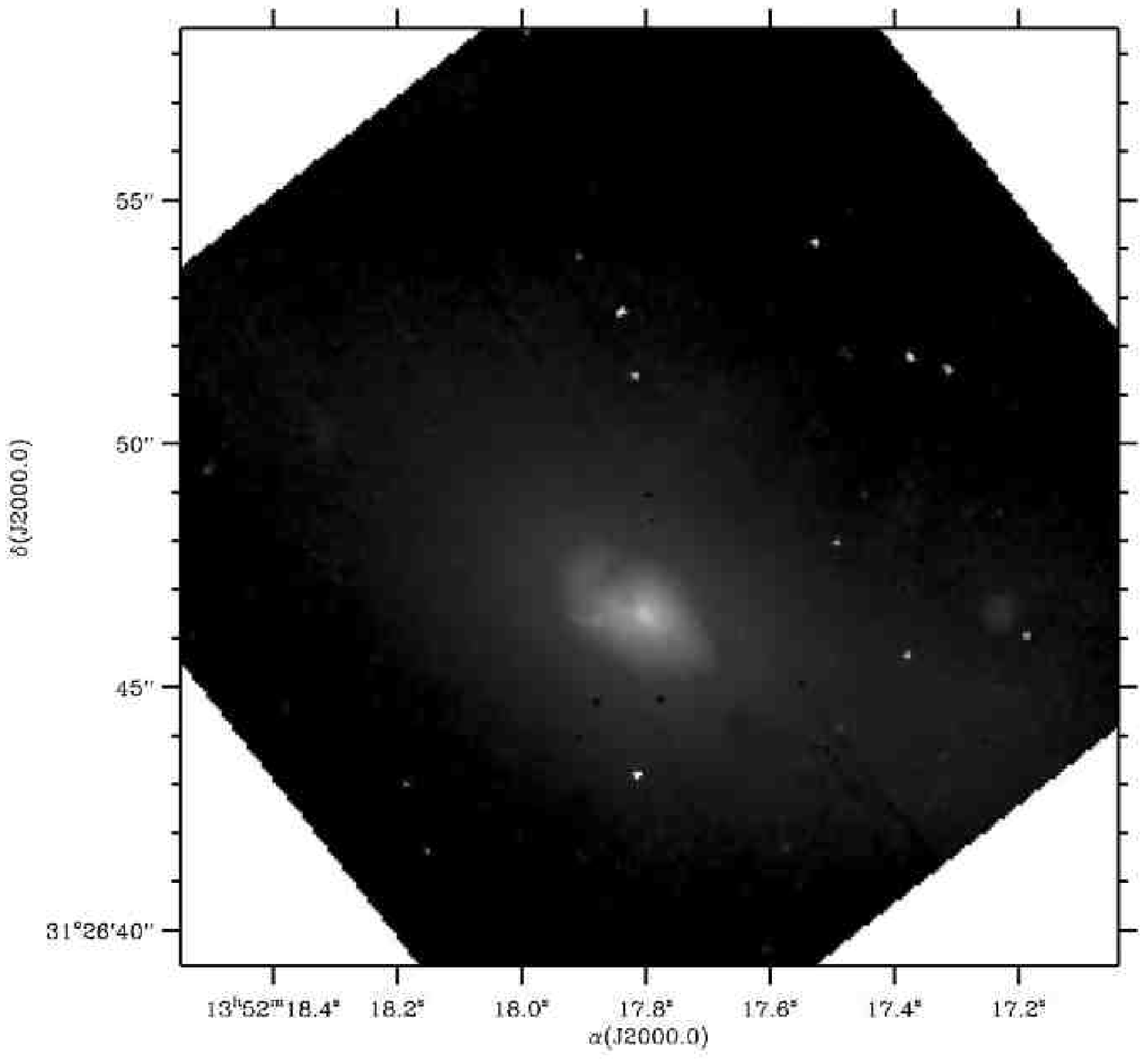}{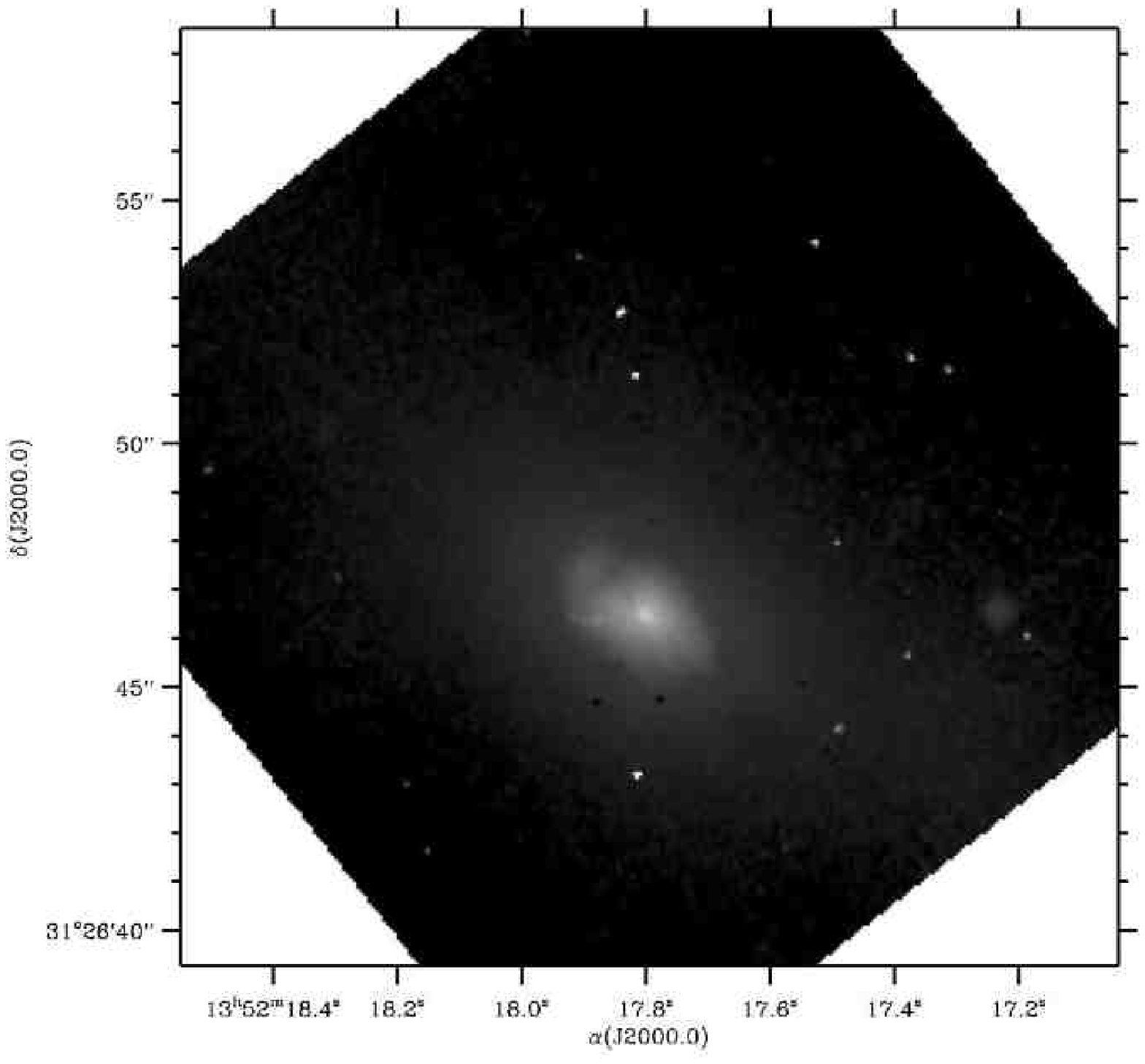}\\
\epsscale{0.35}\plotone{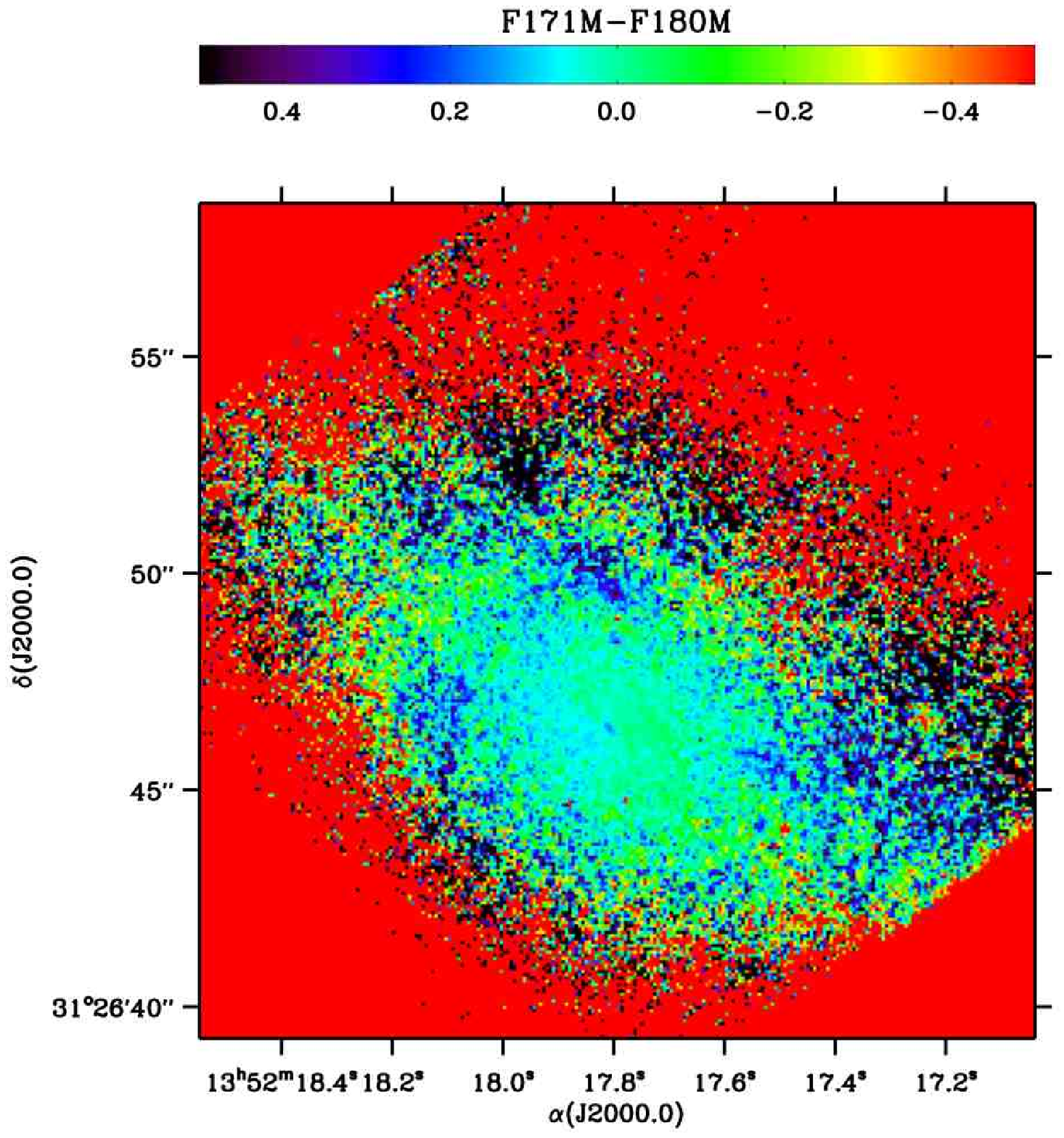}
\caption{\label{fig-narrow} Top left: F171M ([Fe~{\sc ii}]~$\lambda1.644~\mu$m) image of 3C~293. Top right: F180M off-line continuum image. Bottom: F171M$-$F180M emission-line map, in the STMAG magnitude system. Note that the peak here is just off nucleus, as opposed to the peak in the UV flux distribution, which peaks at knot E3.}
\end{figure*}

\subsubsection{Emission lines}
The {\sc f171m} and {\sc f180m} filters provide on and off-line imaging, respectively, of the [Fe~{\sc ii}]$\lambda1.644~\mu$m emission line, which is redshifted close to the center of {\sc f171m} (1.72~$\mu$m at $z=0.04503$). {\sc f180m} is used for off-line continuum measurements, rather than the broad {\sc f160w}, due to the strong colour-gradient in the galaxy~\citep{smithheckman89}. Using the {\sc f171m} and {\sc f180m} images an [Fe~{\sc ii}] image can be constructed (Fig.~\ref{fig-narrow}) by assuming a spectral index of $F_{\nu}\propto\nu^0$ for the underlying continuum (consistent with its flat spectral energy distribution from NED~\footnote{http://nedwww.ipac.caltech.edu/}) and applying a small shift, of the order of 20~mas, to align the images on the galaxy peak and on the E1-E3 knots which are visible in both.

\subsection{Radio Observations}
\label{sec-radio}
\begin{figure*}[htf]
\epsscale{0.7}\plotone{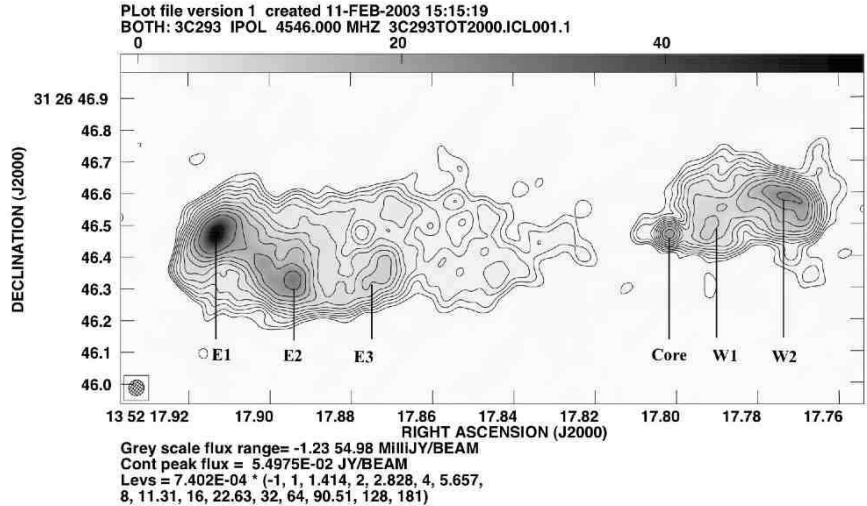}
\caption{\label{fig-newradio} Our new MERLIN 5GHz radio map, generated from a combination of our 	new observations (section~\ref{sec-radio}), and those of Akujor et al. (1996). The image is uniformly weighted with a synthesised beam size of $0.05029\times0.04815\arcsec$ at PA$=-36.01^{\circ}$. It shows only the innermost of the 3 distinct structural scales in the radio emission, comparable to the infrared jet seen in Figs.~\ref{fig-H} and~\ref{fig-K}.}
\end{figure*}

With baselines $\simeq7$ times longer than the VLA A-configuration, the UK Multi-Element Radio-Linked Interferometric Network (MERLIN) array provides resolution almost perfectly matched to HST: FWHM is 46~mas at 6~cm, and 140~mas at 18~cm. MERLIN observations of 3C~293 at 6, 18cm were published by Akujor et al. (1996). We have re-observed at 6cm, the new data including four frequencies in the range 4.5-5.2GHz which gives much better aperture coverage through multi-frequency synthesis, and hence better response to diffuse emission. We have combined our data with those of Akujor et al. to give the 6 cm image presented in Fig.~\ref{fig-newradio}. The new radio data show only the innermost of the three distinct structural scales, the largest of which reaches a size of over 200~kpc. 

\subsection{Alignment of WCS}
The HST (NICMOS and WFPC2) images were aligned using a pair of faint stars at the edge of the field (which only appear in one pointing), using the {\sc iraf} tasks {\sc wcsmap} and {\sc geotran}. No significant improvement was obtained in adding a rotation or geometric distortion, so a simple pixel shift was applied. The STIS and WFPC2 data were similarly aligned, as described in~\cite{allen+02}. 
The HST WCS is only accurate to $\sim1\arcsec$, so we use the coordinate system of the MERLIN radio map to tie the coordinate system to the international coordinate reference frame (ICRF). This was achieved by aligning the 3 knots in the infrared jet to the brightest radio features, which have near-identical separations (Fig.~ \ref{fig-radio-IR}). Note that the radio core also aligns with the center of the infrared host galaxy. We have rotated all images to give North up, and have resampled at $0.04553\arcsec$ per pixel, in order to compare to the WFPC2/PC1 data. Positional accuracy of the matches is in all cases better than a pixel, confirmed by positions of at least three objects in each frame.

\subsection{Colour maps}
In order to produce optical-IR colour maps, the WFPC2 image was smoothed with the NICMOS PSF, generated using TinyTim~\citep{tinytim}, assuming a spectral index of 1.0, to roughly equalize the resolution. This is imperfect as it does not allow for the finite resolution of WFPC2 or the larger pixel size of NIC2, but these two small effects act in opposite directions and most of the structure is resolved. The $R-H$ (computed from {\sc f702w-f160w}) colour map is presented in Fig.~\ref{fig-colmap}.

\section{Observational Results}
\label{sec-res}
Our main observational result is the discovery of an infrared counterpart to the jet seen in the radio. Here we present the near-infrared images and polarimetry, show that the jet is also partially visible in the optical and UV, and demonstrate the synchrotron nature of the emission, through jet SED fitting after subtraction of the host galaxy. IR and radio polarimetry confirms the synchrotron mechanism, and interestingly, no significant [Fe~{\sc ii}] excess is detected. In section~\ref{sec-model} we describe the modelling required to characterise the jet, host galaxy and dust distribution, and in section~\ref{sec-disc} we discuss 3C~293 in the light of these new results. In this paper we adopt the labelling convention of Beswick et al. (2004) to denote the features of the inner, kpc-scale jet (see their Fig.~ 3 and our Figs.~\ref{fig-H} and~\ref{fig-K}). 

\subsection{An optical-infrared-ultraviolet jet counterpart}
\label{sec-images}
\begin{figure*}[htf]
\epsscale{0.75}\plotone{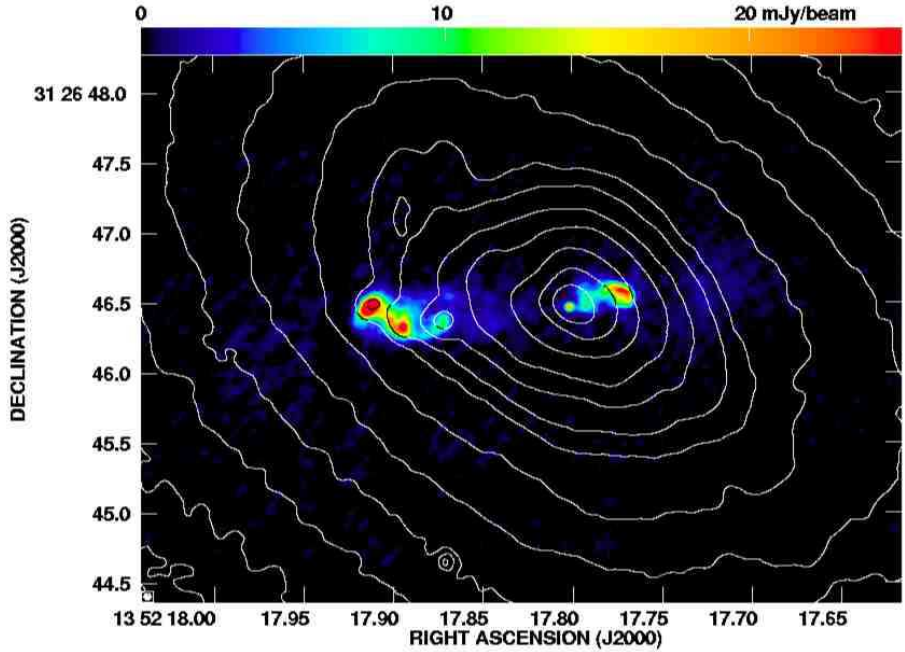}
\caption{\label{fig-radio-IR} $2~\mu$m total intensity ($\approx K$-band) contours overlayed on the 5GHz MERLIN radio map. Knots E1, E2, and E3 are clearly visible in both the radio and at $K$-band. }
\end{figure*}

As seen in Figs.~ \ref{fig-H} and~\ref{fig-K}, the jet is clearly picked out as a string of 3 distinct ``knots'' in the broad-band near-IR images. There is no sign of optical-IR emission in the these images, nor in the model-subtracted residuals (Fig.~\ref{fig-IRpol}), from the counter-jet in the opposite lobe, despite the lower absorption on that side. Moving outward from the nucleus we refer to the jet knots as E3, E2 and E1, following the nomenclature of Fig.~3 in Beswick et al. (2004). All three align well with major features in the radio map (Fig.~ \ref{fig-radio-IR}). Furthermore, the innermost knot, E3, is clearly visible, in hindsight, on both the STIS {\sc f25srf2}, and the $R$-band WFPC2 image, in the latter case as the easternmost of the 2 bright features seen close to the heart of the dust lane (see Fig.~ \ref{fig-optical}). E1 \& E2, are visible as steep-spectrum regions, while E3 does not have a spectrum noticeably different from the surrounding starlight, except for the bright UV point-like emission, which is clearly separable from the non-synchrotron ``background'' of ongoing star-formation (Fig.~\ref{fig-uv}). 

\begin{figure*}[ht]
\epsscale{0.65}\plotone{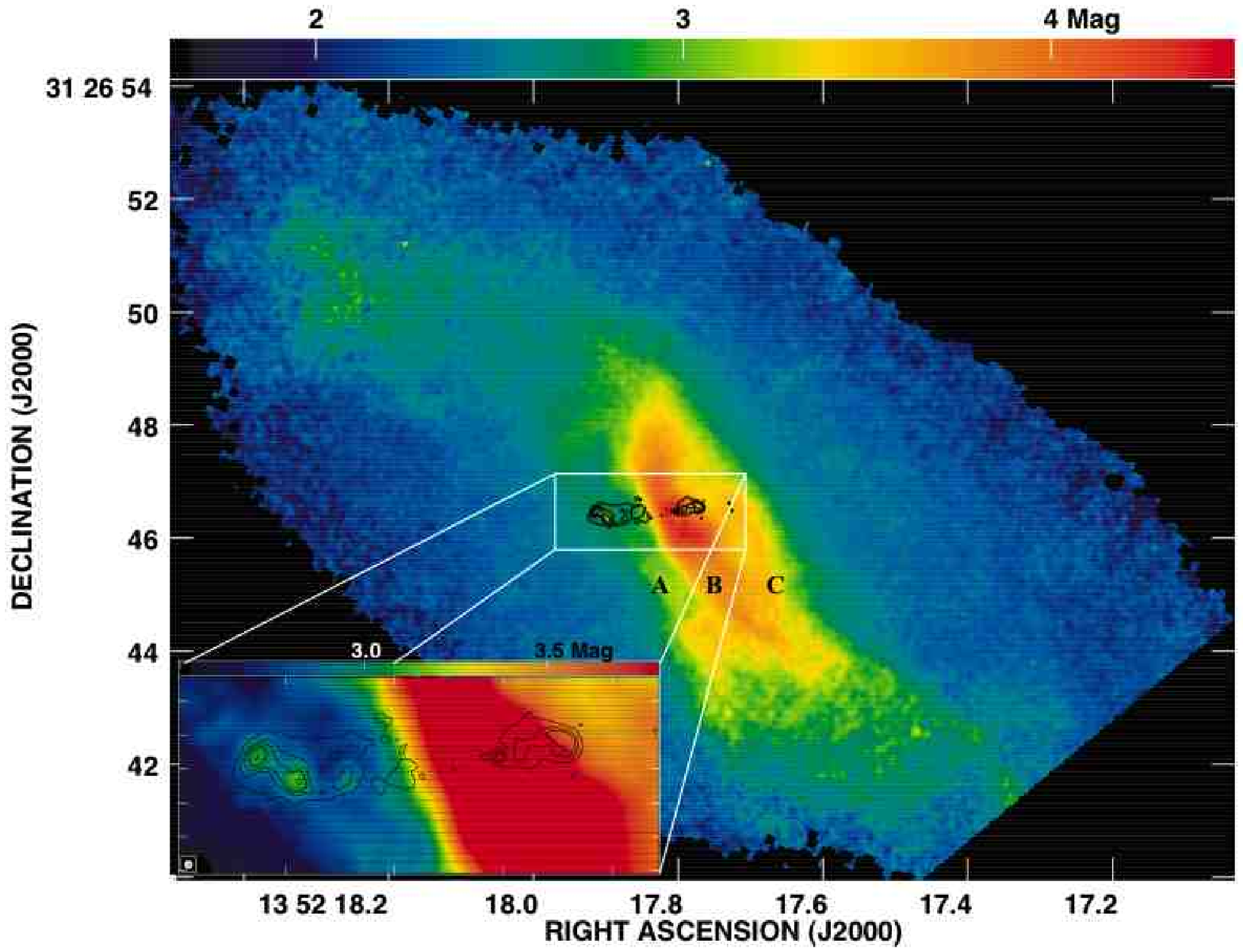}
\caption{\label{fig-colmap} An $R-H$ colour-map, produced from the NICMOS {\sc f160w} and the WFPC2 {\sc f702w} images. The three most prominent dust lanes are labelled A, B and C. C appears to pass behind the bulk of the source, while A and B cross in front of it, consistent with a disk at low inclination. The MERLIN 5GHz contours are overlayed for reference.}
\end{figure*}

No evidence is seen of [Fe~{\sc ii}] emission in the vicinity of the knots, to a limit of about $5\times 10^{-20}$ Wm$^{-2}$. There may be [Fe~{\sc ii}] emission in the central $\sim$200~mas, at a level of about $4\times 10^{-19}$ Wm$^{-2}$. This could, however, be produced by the underlying continuum, which is significantly steeper in the small nuclear region, than in the surrounding area. The limit on the extended [Fe~{\sc ii}] brightness and the nuclear [Fe~{\sc ii}] measurements are both a factor of $\sim$20--50 less than the corresponding values in 3C~305~\citep{jackson+03}.

Significant polarization is detected in the NIR in both the jet and core regions (Fig.~\ref{fig-IRpol}). Clearly most of the light from the core of the galaxy is obscured in the optical by the complex network of dust filaments. The outer jet knots are behind one of these filaments, causing their extinction in the optical, 
and we considered the possibility that the NIR polarization of the outer jet knots could be caused purely by transmission and scattering through this dust. However, the degree of polarization and the agreement with the radio polarization angle~\citep{akujor+96} make this highly unlikely. The outermost knot, E1, shows $\simeq 15\%$ polarized flux at 2~$\mu$m and the B-vector parallel to the jet, and $E$-field orientation at the peak in good agreement with the radio angle~\citep{akujor+96}. The two inner knots have B-vectors perpendicular to the jet, as does the nucleus.

The relative brightnesses of the jet knots are very different at radio and optical wavelengths. The brightest infrared knot, E3, seems rather compact in the WFPC2 image. The corresponding radio feature is faint and clearly resolved at 6cm (compared at the same resolution), so the emission at the two wavelengths is probably not from the same volume. The situation may be similar to that in 3C~273~\citep{bahcall+95} where the narrow optical jet runs through a broader radio lobe. Emission from around the outer two knots (E1, E2) may be faintly visible in the WFPC2 image, but does not appear as two separate peaks: the situation is confused by the filamentary dust lanes, and the apparent emission along the jet may just be starlight visible between two lanes. 
 
 \subsection{An elliptical host galaxy with a single nucleus}
In the near-infrared, we have uncovered the smooth, elliptical structure of the underlying host galaxy, which is heavily obscured in the optical. Even at $2~\mu$m though, it is clear that there is still a small amount of obscuration, and disturbance to the north-east of the nucleus, and north of the jet. Note also the small companion object on the edge of the NICMOS chip in each infrared image, approximately $7\arcsec.5$ to the northeast of the nucleus. This object is also visible on the STIS image (Fig.~\ref{fig-uv}) as a ring of enhanced UV flux in the dusty north-eastern arm of the source, and faintly on the WFPC2 image and $R-H$ colour-map (Figs.~\ref{fig-optical},~\ref{fig-colmap}). 

Fig.~\ref{fig-colmap} shows the $R-H$ colour-map of 3C~293, clearly illustrating the extent and structure of the complex dust lanes that obscure the radio core and the jet, and give the appearance of a complex multi-nucleus host galaxy. The dust lanes appear to be a ring or disk-like structure enshrouding the nucleus, and aligned roughly NE-SW. There is additional, more extended structure that seems to be connected to the SW tidal tail, as well as a number of finer filamentary structures that appear unconnected to the disk, and which may also be tidal in nature. No unresolved nucleus is detected, though the galaxy appears to have a single nucleus. This was also found to be the case for 3C~305. We deduce that any multiple nuclei must fall within the central $<0.1\arcsec$ (i.e. $<100$~pc) if indeed they exist at all. This is contrary to its appearance in the optical image (e.g.~\citealt{martel+99,dekoff+00}) where it appears to have at least two distinct nuclei separated by almost an arcsecond. These are in fact simply the center of the galaxy, and the knot E3.

The complex appearance in the optical is best understood in terms of two foreground gas and dust systems, consistent with the dynamics of the absorption features seen in HI. One of these is rapidly rotating, has a steep velocity gradient (179~km~s$^{-1}$), and must be located close ($\ltsim1$~kpc) to the nucleus. The other, seen as the prominent dust lanes, in optical emission lines, and in narrow H~{\sc i} absorption is located much farther ($\approx 9$~kpc) from the nucleus, and has a correspondingly lower velocity gradient(46~km~s$^{-1}$~arcsec$^{-1}$) - see Beswick et al. (2004), and discussion below.

\section{Modelling}
\label{sec-model}
In order to properly characterise the knot fluxes, it was necessary to undertake detailed two-dimensional modelling of the host galaxy, and apply extinction corrections based on the dust distribution observed in the spectral index map. 

\subsection{Dust and obscuration}
\label{sec-dustmodel}
We have used the multicolour HST images to estimate the total extinction, and generate extinction maps for each of the four UV--NIR images. Firstly, we assume that the outer regions of the galaxy image provide an unobscured view of the stellar population in the host. This is out beyond the edge of the dust lane that is clearly picked out in the colour-map of Fig.~\ref{fig-colmap}.

We note that the UV morphology (Fig.~\ref{fig-uv}) follows that of the absorbing material seen in $R-H$ (Fig.~\ref{fig-colmap}), indicating that the majority of the absorbing material lies in the foreground. We also assume that the overall galaxy colours are dominated by continuum emission, rather than by lines, which is confirmed by the emission line imaging in Fig.~ \ref{fig-narrow}.
$R-K$ is then used to deduce the extinction on a pixel-by-pixel basis, by comparison to the average value $ \langle R-K \rangle$, measured from unobscured regions of the galaxy.
\[(A_{R})_{ij} - (A_{K})_{ij} = (R-K)_{ij} - \langle R-K \rangle \]
The greatest values are found at the center, with up to 3.7 magnitudes of extinction in $R$. Typical values for the dust lanes are $A_{R}\approx 1.5$~magnitudes. We convert between the different bandpasses using the quantities defined in~\citet{mathis90} for their ``diffuse'' dust model ($R_V=3.1$). 
An extinction map, $A_{\lambda}$ was produced for each image at wavelength $\lambda$. This was used to correct the fluxes of the three jet knots (see below).
Column densities are $N(H)\sim3-6\times10^{21}$~cm$^{-2}$ across the jet region, and up to $10^{22}$~cm$^{-2}$ in the densest central regions. Integrating over the central $2\arcsec$ radius, the total mass of dust is found to be $\sim10^{9} M_{\odot}$. Comparing to the $\sim10^{10}M_{\odot}$ neutral gas mass estimates of Evans et al. (1999) yields an approximate gas-to-dust ratio that is somewhat lower than Galactic.

\subsection{The host galaxy}
\label{sec-hostmodel}
In order to accurately characterise the SED of the jet, we first had to model the complex galaxy and absorption ``background.'' This was done using {\sc galfit}~\citep{galfit} to fit the optical and near-infrared images. {\sc Galfit} is a versatile fitting code which allows the user to fit multiple components to a galaxy in an iterative fashion, using a downhill gradient (Levenberg-Marquardt) technique~\citep{numrec}. We present 2 models of the host, which is complex and difficult to parameterise simply. To start with we fit a single-component model to roughly characterise the total flux across the image. Secondly, we treat the host as ``background'' to the jet, and fit it using 3 independent components, in order to produce a smooth background to the jet itself. 

A variance map was generated using the read noise per pixel and the Poissonian shot noise associated with the signal level in each pixel, combined with a static bad-pixel mask. We began fitting with a single S\'{e}rsic profile~\citep{sersic68} component for each band, using a $R-K$ colour-map to construct a dust map to mask out the central regions - both the dust lanes and the jet. We masked the entire core (central $1\arcsec.5$) of the galaxy at this stage, fitting only to the wings. We started the search algorithm with a centering based on the centroid of the object in each image, and using a reasonable estimate of the scale-length, ellipticity, and total magnitude based on photometric measurements made within {\sc iraf}. No constraints were placed on the fit parameters. The results of this single-component model are presented in table~\ref{tab-host-one}. The host is well described by a de Vaucouleurs type profile, and an axial ratio of around 0.6 in each of the three optical-NIR bands.

\begin{figure}[htf]
\centering
\epsscale{0.5}\plottwo{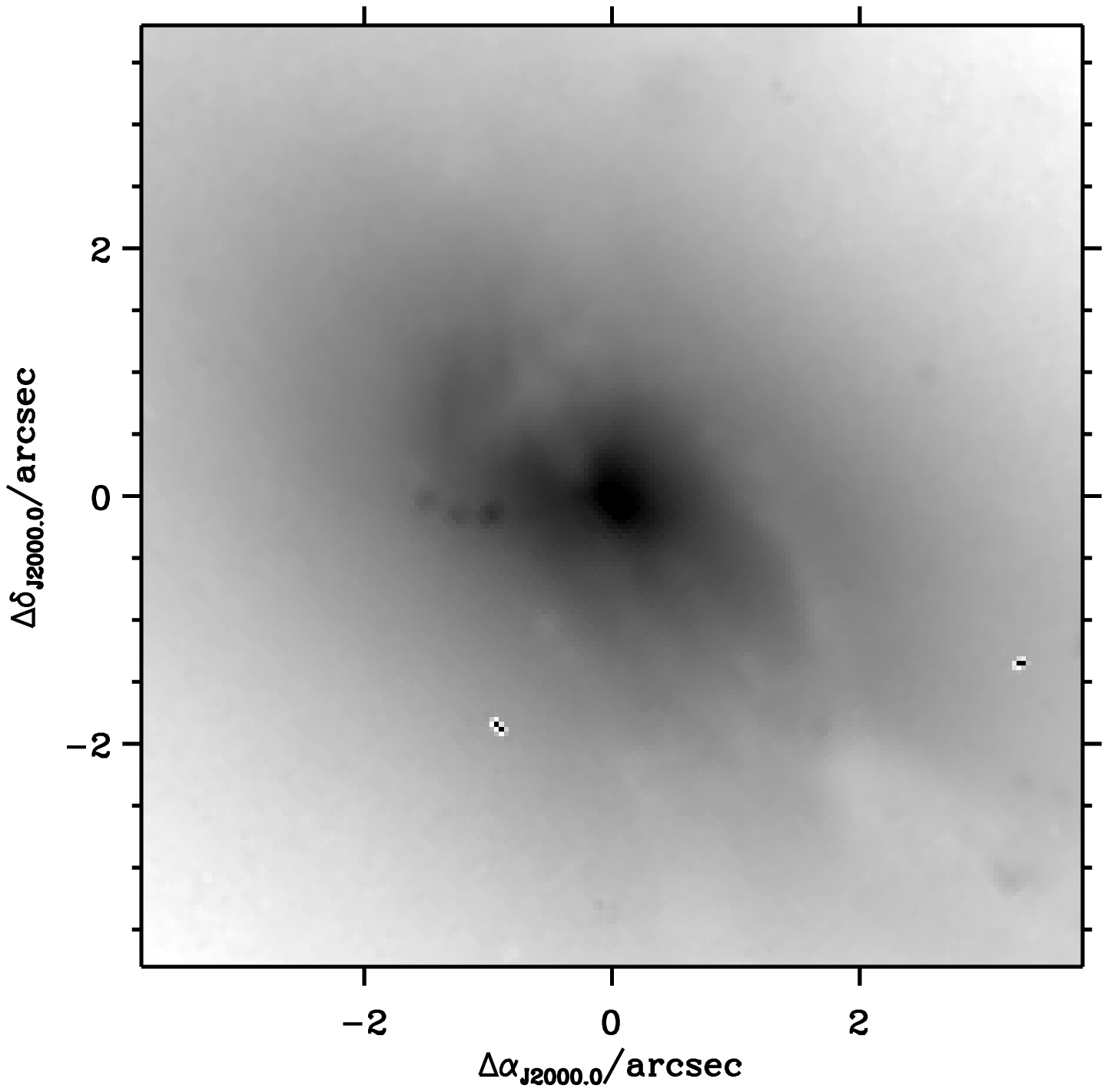}{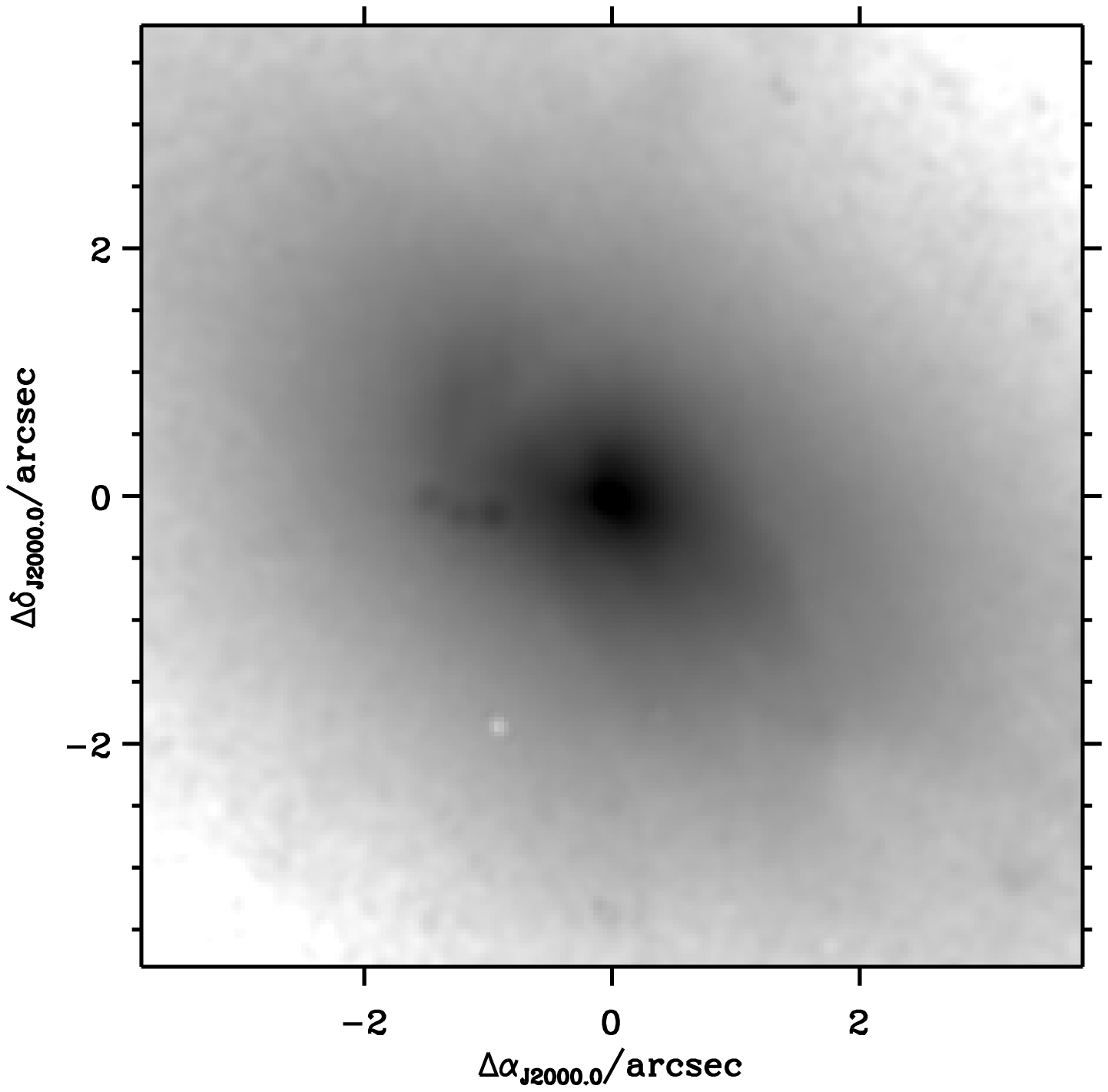}\\
\epsscale{0.5}\plottwo{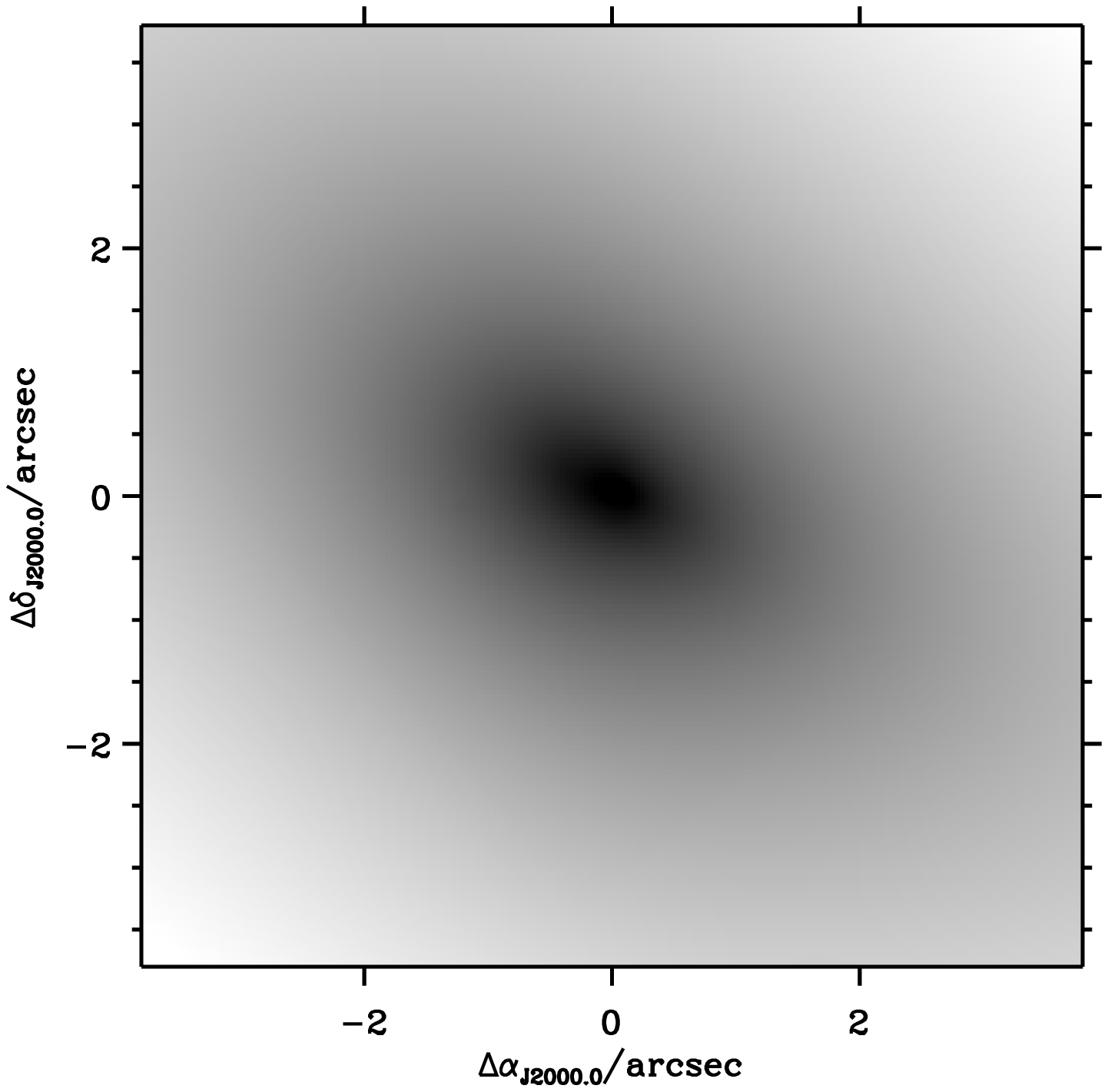}{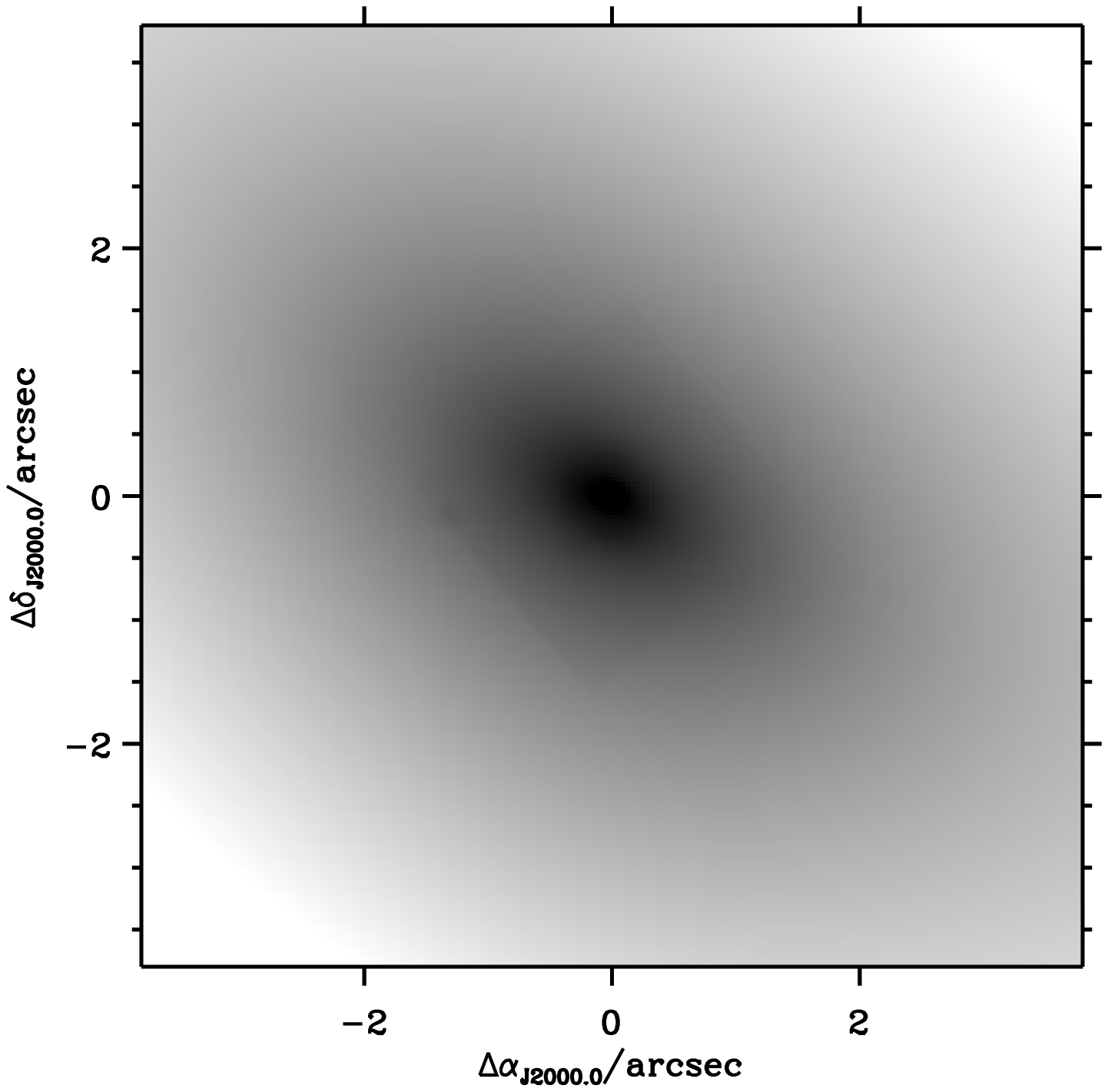}\\
\epsscale{0.5}\plottwo{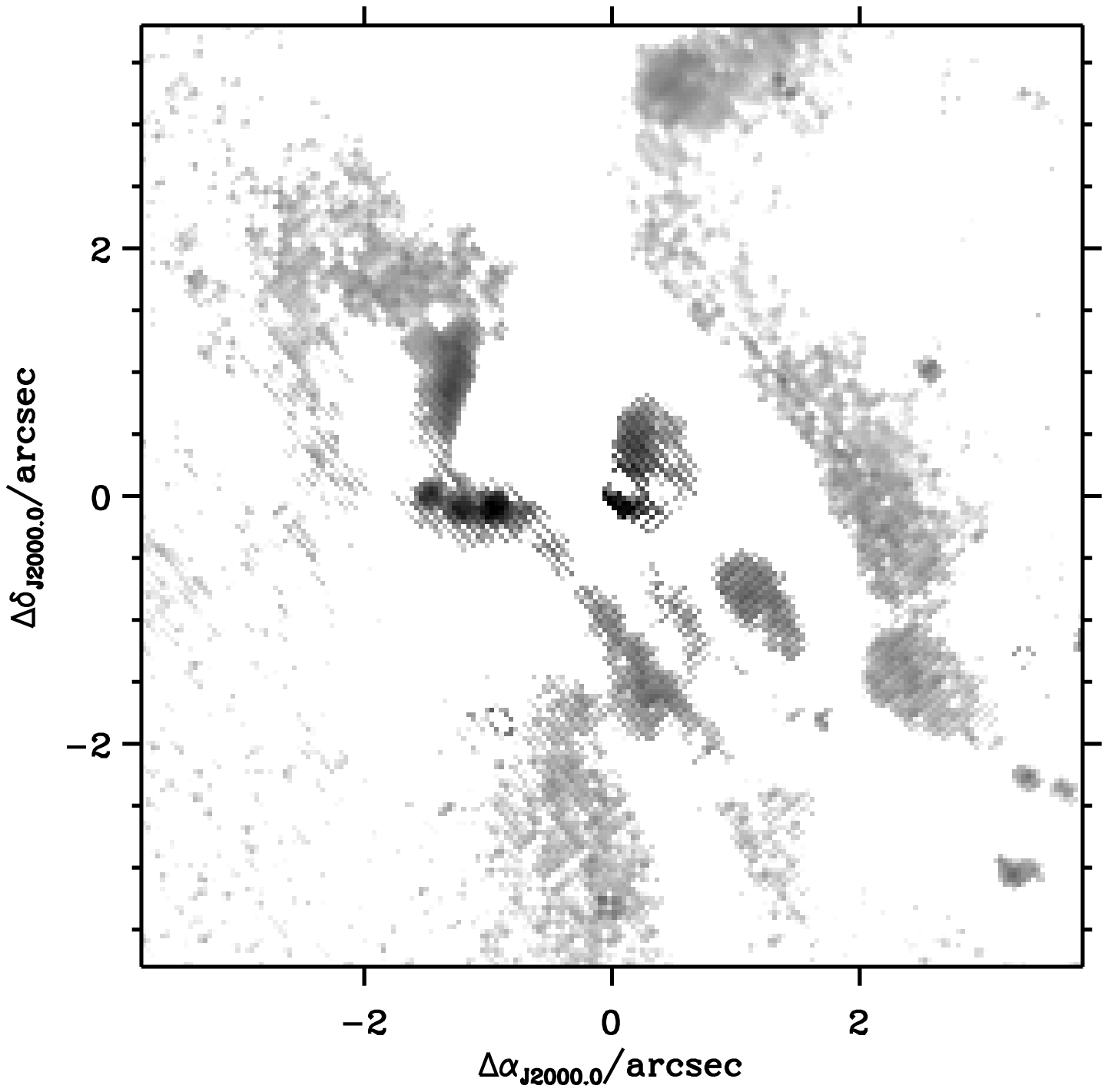}{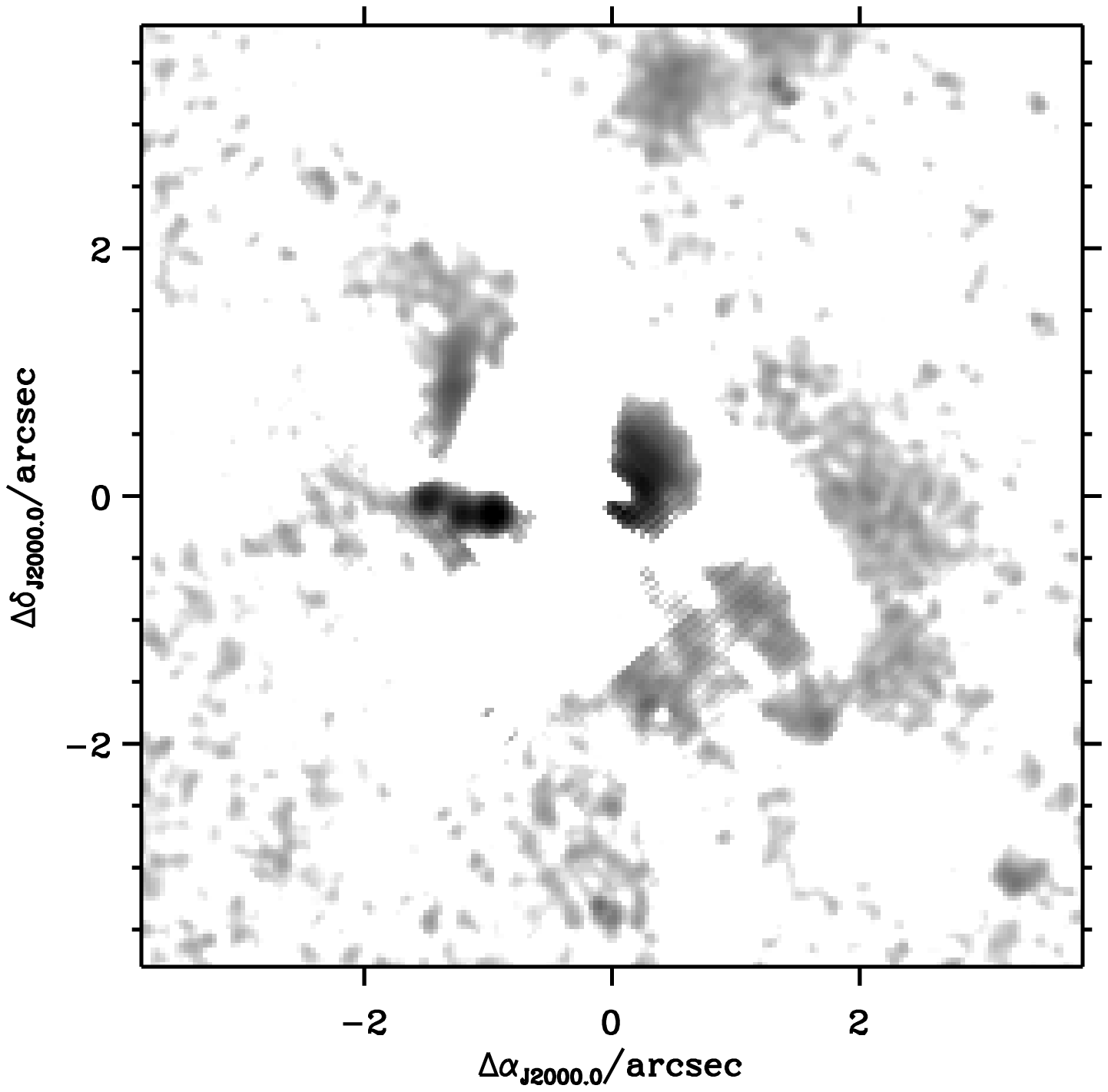}\\
\caption{\label{fig-jet} Modelling the 3C~293 jet at 1.6~$\mu$m (left) and 2.0~$\mu$m (right). Images shown (top-to-bottom) are: The original NIC2 image, the galfit model of host, and the modelling residuals highlighting the jet at a different greyscale to amplify the details. The axis labels are in arcseconds from the nucleus.}
\end{figure}

The dust mask was then relaxed to allow flux from the nuclear region of the galaxy into the fit, while still masking out the dust lanes and jet. We re-defined our dust mask as all regions for which the $H$-band extinction is greater than 0.1mag. The first component was then re-fit to the infrared images (1.6 and 2.0$~\mu$m), with second and third S\'{e}rsic components fit iteratively to the residual maps. The resulting fit provides a ``background'' model for the jet itself - we infer no physical meaning from the model. Due to its complexity, we make no attempt at modelling the $R$-band image in this way - we just fit the infrared images. The jet knots themselves were also fit using a Gaussian profile, which works well for the innermost knots, E2 and E3. The full fits in each band (1.6 and 2.0~$\mu$m) are summarised in tables~\ref{tab-host-multi}, and~\ref{tab-host-multicpt}, and model images and residuals illustrated in Fig.~ \ref{fig-jet}. These model-subtracted residuals are used to characterize the SED of the jet in section~\ref{sec-SED}.

Overall the host galaxy, UGC~8782, is found to be well-fit using three S\'{e}rsic profiles. Whilst the exact structure is too complex to characterise simply, due in part to the dust lane which produces extinction even at $2~\mu$m, we conclude from our single-component fits (which entirely mask out the central region) that the underlying old stellar population forms a large, somewhat eccentric (E4) spheroidal bulge, with a {\em single nucleus}, an unobscured luminosity of close to $L^{\star}$ ($M_{K}^{\star}=-24.17$ converting to $H_{0}=70$~km~s~Mpc$^{-1}$ from the value given by~\citealt{kochanek+01}).

\subsection{The jet SED}
\label{sec-SED}
\begin{figure*}[htf]
\epsscale{0.3}\plotone{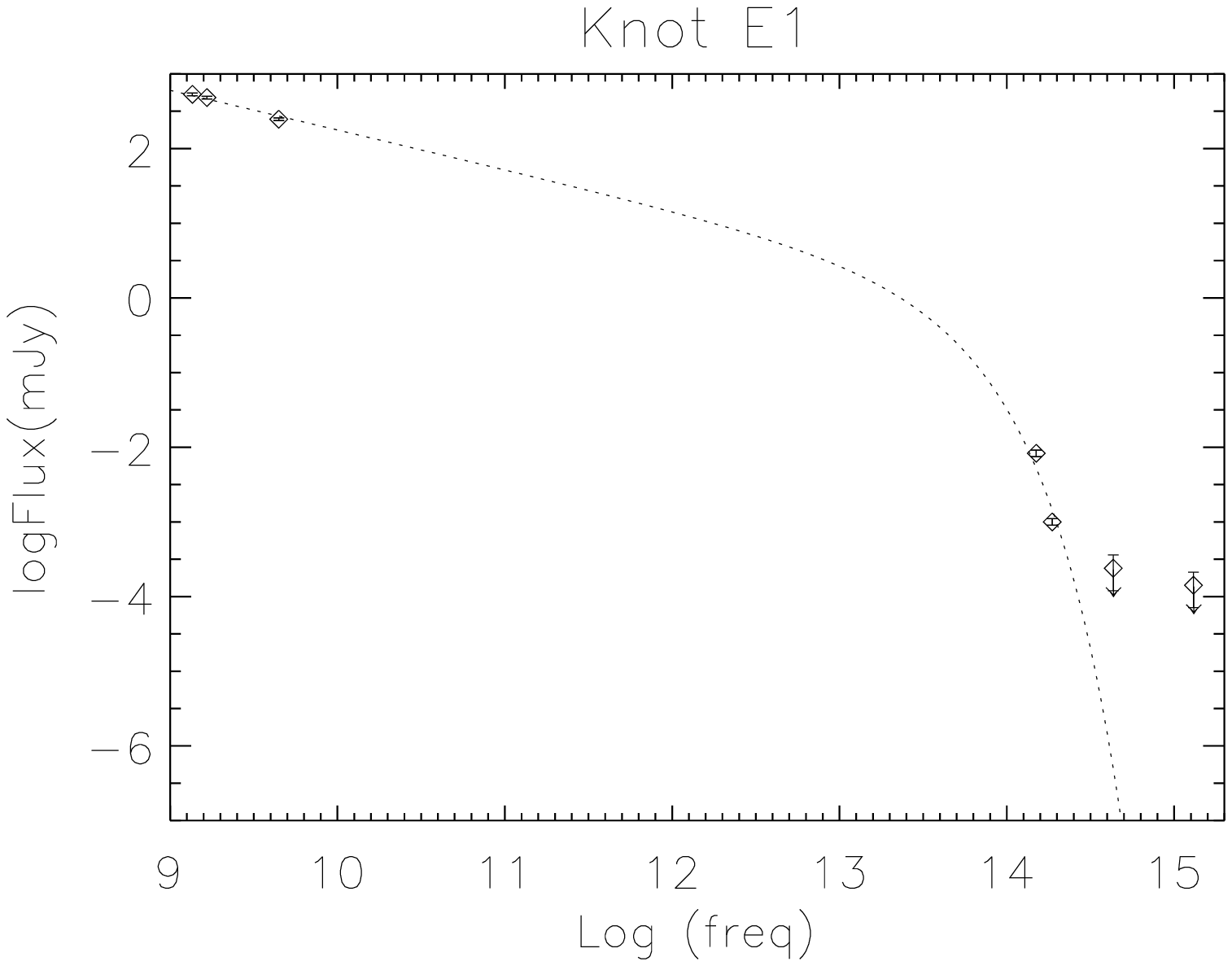}
\epsscale{0.3}\plotone{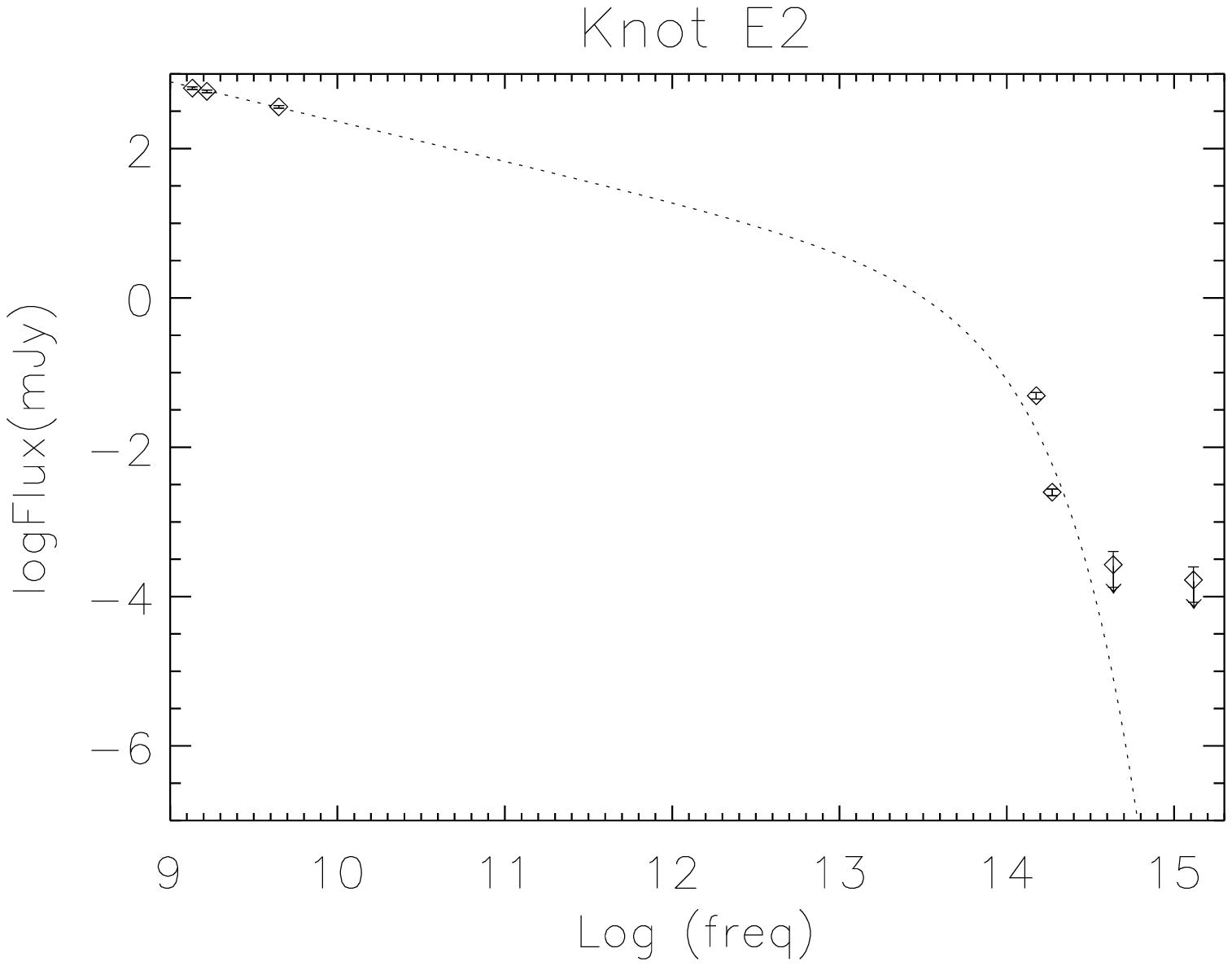}
\epsscale{0.3}\plotone{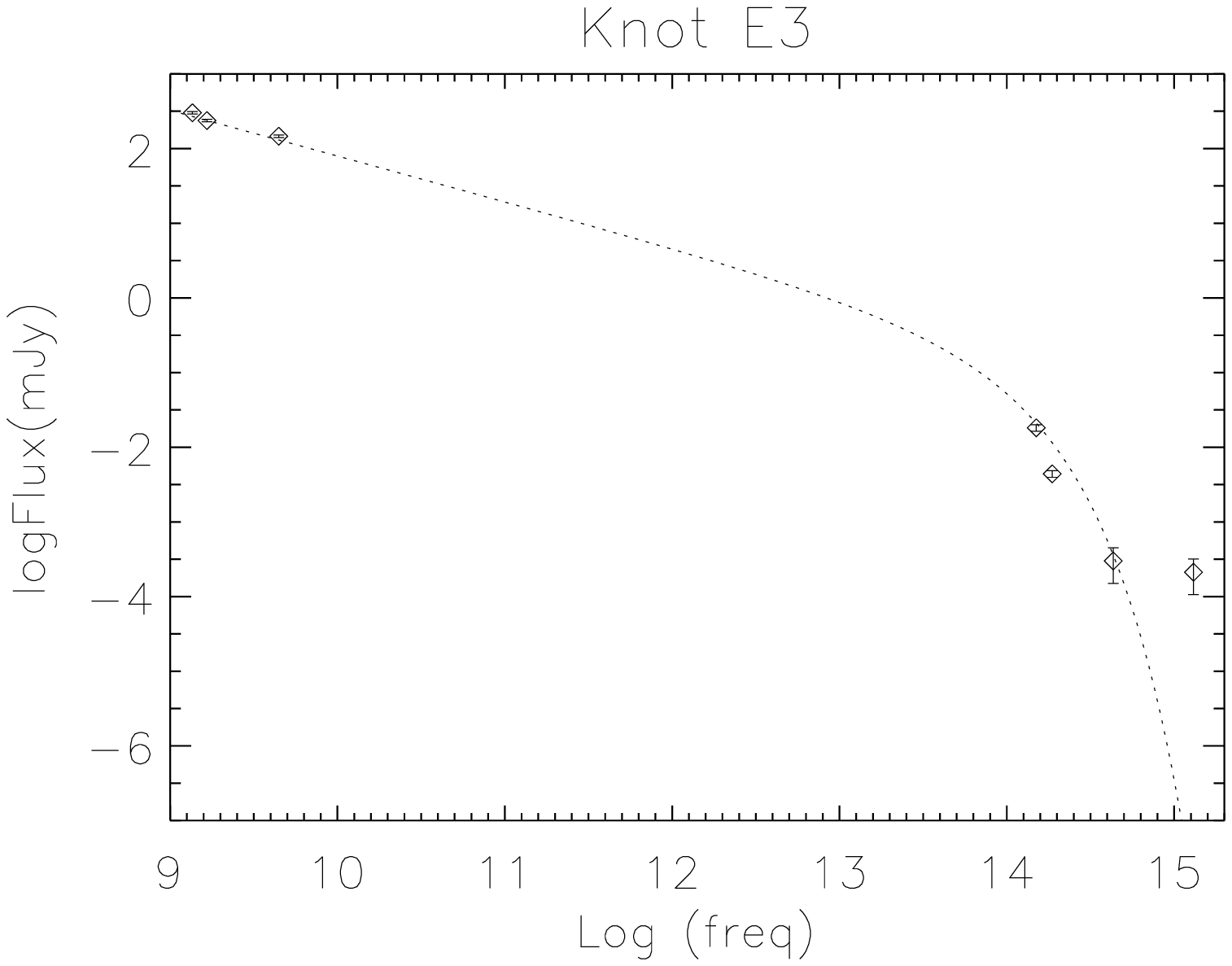}
\caption{\label{fig-SED} SED's of the three knots, E1, E2 and E3, overlayed with synchrotron Jaffe \& Perola (1973) radiation models (see main text). The injection indices and break frequencies are given in table~\ref{tab-SED}.}
\end{figure*}

We extracted spectral energy distributions for each of the three knots visible in the near-IR. To do this we used small photometric boxes to define precisely the physical region for which we extracted flux for each knot. This is particularly important since the morphology of each of the knots is radically different not only in the near-IR as compared to the radio, but also across the radio spectrum itself. For this reason we adopted the simple photometric technique described, rather than fitting of Gaussian profiles. We note that for knots E2 and E3 the NIR fluxes obtained through each method were found to be consistent within the errors. For E1, the Gaussian approximation breaks down. The same regions were used for photometry on the optical and UV images to measure the flux from knot E3. We corrected the fluxes using both the extinction map calculated above, taking the average extinction in each photometric box, as well as the much smaller galactic extinction, $A_{g,B}=0.072$. Fluxes are presented in Fig.~\ref{fig-SED}.

We fit these spectra with synchrotron radiation models using the code of~\cite{leahy91}. In all three cases we assumed a~\citet{jaffeperola73} form for the spectrum. This assumes no particle acceleration or injection within the beam, but does allow for pitch-angle randomization (see~\citealt{leahy91,carilli+91} for a more in-depth description of the Jaffe \& Perola model). The results of these fits are shown in Fig.~ \ref{fig-SED}, with the injection indices and break frequencies given in Table~\ref{tab-SED}. It is clear that in the case E3, the model fails to reproduce the high frequency (in particular ultraviolet) flux. This is discussed further below, but it is clear that some particle re-acceleration is required at this innermost knot.
We also attempted fits using a Kardashev-Pacholczyk~\citep{kardashev62,pacholczyk70} model and a continuous injection model, but these do not cut off quickly enough to match the cutoff seen in the near-IR and optical data, and correspondingly the $\chi^{2}$'s of the fits were significantly worse. The knots clearly show a break in their spectral indices, as has been seen in numerous recent high-resolution studies of optical-IR jets.

\section{Discussion}
\label{sec-disc}
\subsection{3C~293's multi-wavelength jet}
\label{sec-jet}
The infrared jet exhibits good alignment between radio and infrared features allowing us to firmly identify the (rather faint) radio core with the center of the galaxy, and identify the innermost knot on the optical (WFPC2) and UV (STIS) images. This allows us to physically characterise the synchrotron emission across an unusually large range in wavelength, and thus particle energy. Additional evidence that it is indeed synchrotron radiation comes from the high polarization in the infrared, and the similarity between the polarization patterns in the infrared and radio. The presence of such high-energy synchrotron radiation suggests that there may be an {\em in situ} acceleration mechanism at work, since the radiative timescales for synchrotron emitting electrons in the near-infrared is $\sim10^{3}$~yr, and even shorter for the optical and UV ($\tau\propto\nu^{-1/2}$). This appears to be the case for the innermost knot, E3, which lies at a projected distance of almost 2~kly from the nucleus, and features significant high frequency flux that is unexplained by the basic Jaffe-Perola model that fits the other two knots well.
 
\subsubsection{Polarization}
\label{sec-jetpol}
The multi-wavelength radio polarization images of Akujor et al. (1996) show a good example of the asymmetric depolarization typical in one-sided jets - the ``Laing-Garrington effect''~\citep{laing88,garrington+88}. The overall depolarization is strong, but it is significantly stronger on the western side of the inner jet structure, implying that the inner western jet is situated behind more obscuring material. There are clearly differences in the morphology of the polarized emission in the near-IR (Fig.~ \ref{fig-IRpol}) and the radio (Akujor et al. 1996). First of all, the polarizations are slightly higher in the infrared compared to the radio (average 15\% in the near-IR as compared to a {\em maximum} of 13\% in the radio~\citealt{akujor+96}). Little NIR polarization is observed on the western side, although the core is slightly polarized. However, this is consistent with dichroic absorption, as the B-vectors are found to be perpendicular to the jet direction. The maximum fractional polarizations in the radio are on the jet's northern side, while in the near-IR we observe higher polarizations on the jet's southern side. In the near-IR, the two inner knots (E2, E3) display a pattern of lower fractional polarization at knot maxima, with higher values upstream and downstream of maxima. Such a pattern is not observed in the radio, where the polarization minima are generally downstream of the knot maxima. This pattern is similar to what has been observed in the knots of M87~\citep{perlman+99}, but runs counter to what has been seen in the jet of 3C~273~\citep{thomson+93}.

The pattern displayed by the polarization B-vectors (plotted in Fig.~\ref{fig-IRpol}) is more complex. In the near-IR we see generally near-perpendicular (to the jet) magnetic field vectors throughout components E2 (PA$_{B}=160^{\circ}\pm 10^{\circ}$) and E3 (PA$_{B}=130^{\circ}\pm10^{\circ}$).
However near the flux maximum of E3 (but not E2) the vectors' PA rotates by up to 40 degrees. Such a rotation is not seen in the radio (Akujor et al. 1996), again similar to what is seen in several knots in the M87 jet~\citep{perlman+99}, where the pattern was interpreted in terms of a stratified jet, with high-energy particles concentrated near the jet center. This again appears to be the case with E3 having the clearest evidence for particle acceleration.
Component E1 however, displays in the near-IR, magnetic field vectors on the northern side that are almost parallel to the jet direction (PA$_{B}=100^{\circ}\pm 10^{\circ}$), i.e., a change in PA of about 60$^{\circ}$ from what is observed in E2. No such abrupt changes are seen in the radio.

\subsubsection{Ionisation}
3C~293 is clearly a gas-rich system with strong ionisation~\citep{vanbreugel+84}. Infra-red [Fe~{\sc ii}] is potentially important as a diagnostic of emission from shocks because it can be enhanced by grain destruction in shocks~\citep{moorwoodoliva88,forbesward93}. It can also be modelled in pure-photoionisation models~\citep{simpson+96,oliva+01}. Its absence in the infrared knots is further evidence that they are part of a synchrotron jet. More general conclusions about excitation mechanisms are difficult to draw given the lack of detailed high-resolution emission-line images in other lines, although to zeroth order the non-detection indicates that the dominant line-excitation process is likely to be photoionisation.

\subsubsection{SED}
\label{sec-jetsed}
The SED's for the outer two of the three jet components, E1 and E2, are fit well using a Jaffe-Perola model~\citep{jaffeperola73}, as can be seen from Fig.~ \ref{fig-SED} and Table~\ref{tab-SED}. Note that the two knots have remarkably similar break frequencies, $\nu_{b} \approx 4 \times 10^{13} {\rm Hz}$, despite their different morphologies, and distances from the nucleus varying by $\sim$kly.
Behind this apparent similarity, however, there are some differences, with knot E2 having a significantly flatter injection index, and knot E1 having by far the steepest spectral index in the infrared. However, these are considerably smaller spectral changes than seen in, for example, M87~\citep{perlman+01,perlmanwilson05} or 3C~273~\citep{jester+01,jester+05}. 
This is in accord with the much smaller changes seen in the $2~\mu$m  polarization map (Fig.~ \ref{fig-IRpol}), which show relatively constant  polarization and direction of magnetic field throughout the three knots. E3 is most like regions in the M87 jet, with active particle acceleration clearly underway, and a similar polisation pattern. 

Our fits to the spectral data for knot E3 show that there is a clear need for an additional emission component in the ultraviolet.  This component may also contribute a significant (or even dominant) fraction of the emission in the {\sc f702w} image.  The origin of this component is unclear.  One possibility is that it is synchrotron emission from a small (high-energy tail) component of the electrons that radiate at lower frequencies.  A second possibility is that it is inverse-Compton emission from low-energy ($\gamma<100$) electrons.  Optical inverse-Compton emission has never before been detected in an optical jet source, however, given the similarity between the {\sc f702w}-STIS and radio spectral indices we cannot rule out this hypothesis. A final possibility is that the excess UV flux is produced by star-formation activity that has been induced by the jet. Scheduled Chandra observations may help discriminate between these three possibilities.

The combination of fixed break frequency and small changes in the PA of the magnetic field suggest a distributed particle acceleration mechanism such as in 3C~273, rather than one concentrated in discrete sites like in M87. The X-ray flux and spectral index maps seen in M87 also require a distributed component of particle acceleration along with the knot-specific mode~\citep{perlmanwilson05}. If there were discrete locations with shock acceleration we would see areas in which the  polarization vectors become perpendicular. We note that in spite of the lower resolution, an M87-like transverse B-field would be visible in the present data if it existed in 3C~293.

\subsubsection{Orientation}
\label{sec-jetor}
Perhaps the most interesting feature of the new jet is that it is seen on the eastern side of the galaxy, which is the more complex, being heavily obscured by the heavy dust lanes, and a disk-like feature which pass in front of the jet. \citet{emonts+05b} have localised the very high-velocity blue-shifted warm gas outflow to the eastern side of the source, clearly indicating that this material is approaching us. Secondly, the depolarization observed in the radio continuum means that more ionised gas is present along line of sight towards the western jet than the eastern one, implying that it is receding from us. Finally, the infrared jet's asymmetry confirms that as in every other known case, the optical-IR jet is on the approaching side, implying that it is probably significantly beamed. 

The severe misalignment of the jet and disk-like dust-lanes seen in the optical, NIR, and UV has been hinted at before, but becomes extremely clear from the new data and analysis presented here, which clearly demonstrates that it is the western side of the galaxy that is the nearest (lying to the observer's side of the disk-like dust structure), and yet the eastern jet that is approaching. The dust disk itself aligns roughly with the galaxy, although some of the dust lanes appear to cut across (see section~\ref{sec-dust} below). 


The large-scale radio lobes, however, are aligned more-or-less perpendicular (in projection) to the major axis of the disk-like dust lanes, and thus with the major axis of the galaxy itself. According to Beswick et al. (2004) and Evans et al. (1999), it is the western lobe that is approaching, but the evidence (based on asymmetry arguments) is poor. Asymmetries in the radio lobes are most likely due to the clumpiness of the environment.

\subsection{Gas and dust absorption}
\label{sec-dust}
Beswick et al. (2004) showed that the H~{\sc i} absorption in 3C~293 is split into two dynamically different, and spatially resolved systems. One of these is narrow ($<50$~km~s$^{-1}$, centered at 13500~km~s$^{-1}$ with a weak velocity gradient $\sim46$~km~s$^{-1}$~arcsec$^{-1}$) and identified with the narrow [O~{\sc II}]/[O~{\sc iii}] emission seen in the optical spectroscopy of~\citet{vanbreugel+84}. The optical emission lines are observed with a velocity gradient of $\sim40$~km~s$^{-1}$~arcsec$^{-1}$ out to a radius of $\sim10$~kpc, where we can no longer see the H~{\sc i} lines due to the lack of background continuum for them to absorb. The second is much broader ($>60$~km~s$^{-1}$, seen only against the western half and with a strong velocity gradient of $\sim179$~km~s$^{-1}$~arcsec$^{-1}$), and is interpreted either in terms of a rotating system, or in terms of two further gas layers at different velocities and distances along the line of sight. To this a third, ``very broad'' ($\Delta v_{\mathrm FWZI}\sim1000$~km~s$^{-1}$) H~{\sc i} absorption component has been added by~\citet{morganti+03}, which has recently been interpreted by~\citet{emonts+05b} as ionised and neutral gas in outflow from the eastern (jet-side) of the source.

The obscuration in the optical appears to be due to two components, similar in nature to the two main components detected in H~{\sc i} absorption. We see large scale, rather nebulous tidal features creating the dust lanes that criss-cross the nucleus of the galaxy, and a more coherent, possibly disk-like feature in the central region. The situation is complicated by the superposition which creates the appearance that the two systems run into one another.

\begin{figure*}[htf]
\epsscale{0.6}\plotone{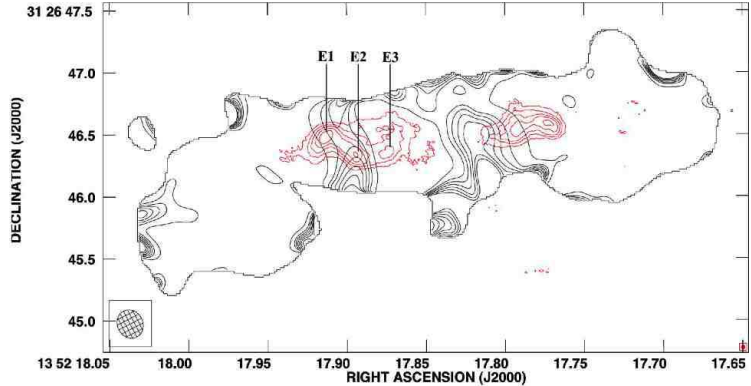}
\caption{\label{fig-optd} Optical depth moment 0 of the narrow H~{\sc i} absorption component, overlayed onto the outline of the radio structure seen with MERLIN at 1.359GHz. The positions of the H~{\sc i} absorption can clearly be seen to coincide with the positions of the dust lanes seen in the optical image and optical-infrared colour map (Fig.~\ref{fig-colmap}). VLBI continuum is shown in red for scale, with contours at 1, 2, 4, 8, 16, 32 and 64$\times$1.3mJy/beam.}
\end{figure*}

We interpret the disk-like feature as tidally disrupted material which may encircle the nucleus at $\ltsim10$~kpc radius, producing the appearance of a much more coherent disk due to its small inclination, and proximity to the line-of-sight to the nucleus of the galaxy on the near side. We argue that the narrow H~{\sc i} feature seen obscuring the region just east of the core in Fig.~ \ref{fig-optd} is an extension of the major southwestern tidal feature (seen in Fig.~\ref{fig-colmap}) which runs northeast across the nucleus, and appears to merge with material that encircles the nucleus. This feature is detected in H~{\sc i} absorption, with a constant velocity gradient across the source of 46~km~s$^{-1}$~arcsec$^{-1}$, and in optical emission lines with consistent velocity gradient out to a 10~kpc radius, where we no longer see the H~{\sc i} absorption due to a lack of background radio continuum. By directly comparing Fig.~\ref{fig-colmap} with Fig.~\ref{fig-optd} we can clearly identify two of the three most prominent optical dust lanes with the 2 major narrow H~{\sc i} absorption features; one crossing E1/E2 (``A'' in Fig.~\ref{fig-colmap}), and the other (``B'') crossing just east of the core. The third prominent dust lane, ``C,'' most likely passes behind the radio source (and thus also behind the bulk of the galaxy) as it is not observed in absorption. The velocity gradient observed in the narrow absorption-line material is consistent with a ring having a radius of $\approx 8$~kpc (Beswick et al 2004), and implies that the eastern side is rotating away from us. This ring-like structure, has an apparent axial ratio $0.3<b/a<0.6$, implying an inclination angle $\sim20-35^{\circ}$ if it is circular. This would be consistent with the appearance of the jet behind the ring (Note that E3 appears just east of the eastern limb), with an inclination angle of $\sim55^{\circ}$ (Beswick et al. 2004).

Regardless of whether the optical obscuration is caused entirely by tidal gas at a large distance (several kpc) from the nucleus, or partially by a circumnuclear disk of a few hundred parsec radius, it is clear that the NIR jet is observed in between two regions of obscuration, with an obscuring tail of material crossing a line of sight that passes in between E2 and E3 (Figs.~ \ref{fig-colmap},~ \ref{fig-optd}).
The Western feature (the core plus counter-jet) is seen through a large amount of obscuring material - through the top of the gaseous disk/ring. The complex of H~{\sc i} absorption seen against the western arm fits in with a picture of obscuration by faster-moving material closer to the nucleus of the galaxy. This would seem to support our interpretation of the features seen in Fig.~\ref{fig-colmap} as a few kpc circumnuclear disk or ring, plus fainter foreground tidal material at a greater distance ($\sim10$~kpc) from the nucleus. Note that~\cite{noel-storr-phd} found that two distinct kinematic systems are common in radio galaxies.

\subsection{Host galaxy}
\label{sec-host}
It is clear from our new infrared images that the galaxy possesses a single nucleus (or any multiple nuclei are within the central $\ltsim100$~pc). Indeed it is found that most, if not all ``merger-like'' radio galaxies have single, well-defined nuclei. This has interesting ramifications for merger-driven AGN scenarios, suggesting that:
\begin{enumerate}
\item They are mergers of unequal-sized galaxies. 
\item The mergers are quite ``old,'' and the nuclei have merged.
\item The mergers are very ``young,'' and the molecular gas came from the companion galaxy (see~\citealt{hernquistmihos94,mihoshernquist96}.
\end{enumerate}

The three companions identified by \citet{emonts+05a} are too distant ($\gtsim100$~kpc) to be considered as merger remnants. There are thus two potential nearby merger remnants; the small companion object found $37\arcsec$ to the SW, and an even smaller remnant seen on the NE edge of the NIC2 chip in our NIR images (Figs.~\ref{fig-H},~\ref{fig-K}), some $7\arcsec.5$ from the nucleus. The latter is found to have a ring of UV flux surrounding it on the STIS image (Fig.~\ref{fig-uv}), and is clearly detected at $H$ and $K$-band, less clearly in $R$ as it is obscured by one of the major dust lanes. However, after correction for extinction it has an $R-H$ colour of 2.5, broadly consistent with an old stellar population, but surrounded by a ring-like feature in the UV indicating active star formation.
It seems unlikely that either small companion is the source of all the disturbance in the main galaxy which contains $O(10^{10})\mathrm{M}_{\odot}$ of molecular gas (Evans et al. 1999), as well as extensive regions of ionised and atomic gas, which appear to be in steady rotation (van Breugel et al. 1984). If either of these sources is the remnant of the merged companion galaxy, then the precursor must have been a very late-type spiral colliding at very low impact parameter. This would allow the companion galaxy to be stripped of its gas and dust extremely efficiently, leaving only a small bulge as a remnant. However, it is possible that the companions (especially the southwestern one) are examples of dwarf galaxies forming within a tidal tail~\citep{barneshernquist92a}.
Cygnus A has also recently been shown to have a compact red companion source, posited by~\citet{canalizo+03} as a stripped stellar component from a minor merger, and by~\citet{bellamytadhunter04} as an infalling molecular gas cloud.

\cite{tadhunter+05} find that the young stellar population (YSP) in 3C~293 is fairly old ($\sim$Gyr) and massive. However the age is poorly constrained with any YSP with an age greater than 0.075~Gyr producing an acceptable fit, with significant degeneracies between reddening and age of the YSP. For example, it is possible to fit a relatively young YSP ($<0.3$~Gyr) with large reddening [$E(B-V)>1.0$] or a relatively old YSP ($1.0-2.5$~Gyr) with moderate reddening [$0.3<E(B-V)<0.5$]. They break this degeneracy using the Ca~{\sc ii} K line strength, which shows that the YSP is an intermediate age population, though with significantly younger ages on the eastern side, where the NIR jet is detected (and where much of the UV flux is concentrated). Tadhunter et al. conclude that 3C~293 would have appeared as a LIRG or ULIRG if observed within 10~Myr of the start of the starburst. 

Finally, we note that given the large quantity of molecular gas estimated by Evans et al. 1999 ($\sim10^{10}M_{\odot}$, several times that of a giant spiral like the Milky Way), the FIR luminosity is surprisingly low ($L_{8-1000~\mu\mathrm{m}}=33\times10^{9}L_{\odot}$) compared with the LIRGS 
($L_{8-1000~\mu\mathrm{m}}>10^{11}L_{\odot}$), which have comparable gas masses. This suggests that either the merger is {\em very} recent, and the star-formation has not had time to get underway properly, or that something in the galaxy is suppressing the star-formation rate.
It is also possible that the CO mass estimate is incorrect, as it is based on an ``X-factor,'' $X_{CO}=2.5\times 10^{20}$, deduced from calibrations in the Milky Way, where conditions are quite different. An order of magnitude smaller CO mass would fit the accretion of a dwarf galaxy, which is more consistent with the appearance of the candidate merger remnants. The star formation that we can see is on the edges of the dust lanes in the UV, and as such is a poor indicator of the true star-formation rate~\citep{surace+00}. It is possible that the strong outflow of neutral gas detected by~\citet{morganti+03}, and~\citet{emonts+05b} is responsible for such a suppression, although at present it is impossible to constrain the cause of this outflow. Higher-resolution CO maps would resolve the morphology of the molecular gas better, and be useful in placing constraints on the star-formation, and how it is being influenced by the jet.

\section{Conclusions}
We have discovered an infrared counterpart to the inner radio jet in 3C~293. Three knots are seen at 1.6~$\mu$m and 2~$\mu$m, corresponding to the three brightest features in the eastern radio jet. The jet and core flux is strongly  polarized at 2~$\mu$m. We have shown that the innermost of these three knots is also visible on archival WFPC2 ({\sc f702w}) and STIS ({\sc f25srf2}) images of the object. This has allowed us to characterise the jet emission as synchrotron, over an unusually wide spectral range. Particle acceleration is evident at the innermost knot, but it is unclear what the acceleration mechanism is. High-resolution X-ray data are the natural next step in properly constraining the physics of the emitting material, although we note that even Chandra's $\gtsim 0\arcsec.5$ resolution will barely resolve the jet into the constituent components studied here. No [Fe~{\sc ii}] emission is detected around the jet, reinforcing the synchrotron interpretation, and implying that shock-driven ionisation is unimportant. This suggests, to zeroth order, that photoionisation is the dominant line excitation mechanism in this source.

3C~293 is a clearly a post-merger system, but it is evident from our 2~$\mu$m image that 3C~293 has, like most merger-like radio galaxies, only a single nucleus. The FIR luminosity of 3C~293 is surprisingly low given the large quantity of molecular gas clearly present, and we argue that unless the merger is very recent, something is suppressing the star-formation rate. It is possible that a jet-driven outflow of neutral gas could do this, yielding the impressive outflow of neutral material seen in H~{\sc i} absorption~\citep{morganti+03,emonts+05a,emonts+05b}. The UV gives a distorted view of star-formation in such post-merger sources, and higher spatial resolution CO observations (e.g. CARMA which will give $\sim 0\arcsec.3$ resolution at 3~mm) are required in order to explore the morphology of the neutral gas, and explain the low star-formation rate.

The large quantities of gas and dust now enshrouding UGC8782 are consistent with a large late-type spiral galaxy being the merged object. There are two obvious candidates for merger remnants - the small companion $37\arcsec$ SW, and a still smaller companion $7\arcsec.5$ NE. The nature of these objects is potentially critical to our understanding of the events that led to the current phase of activity. We note that the SW companion is exceptionally blue, appearing only very faintly in $K$-band. Thus we conclude that its originator must have been a dwarf, or an extremely late-type galaxy with a very small bulge, or else that this object is unrelated to the merger, and has formed post-merger, from the material in the tidal tails. The newly-detected NE object on the edge of our infrared images has optical-IR colours consistent with an old stellar population

We see two absorption systems, consistent with those observed in HI. Firstly, there is a dense molecular gas complex (possibly a ring or disk, a few hundred pc in diameter), and with large velocity dispersion. This provides the broader H~{\sc i} absorption with the high velocity gradient (close to nucleus of galaxy), as well as the CO emission and absorption.
The second system is at larger ($\ltsim 10$~kpc) radius, and is observed as the dust lanes, and in the narrower H~{\sc i} absorption which pass in front of the bulk of the galaxy. The smaller velocity gradient points to a larger radius. The NW side of the galaxy is on the ``near'' side of the disk, with the SE partially hidden beneath the disk of gas and dust. The eastern jet is seen emerging from under the nearside edge of the disk. The kpc-scale jet is oriented quite differently to the large-scale radio structure, but the rotation in both absorption systems is in the correct sense to explain the change in radio axis, as does the alignment of the galaxy from jet buoyancy effects.



\section{Acknowledgements}
Based on observations with the NASA/ESA Hubble Space Telescope, obtained at the Space Telescope Science Institute, which is operated by the Association of Universities for Research in Astronomy, Inc. (AURA), under NASA contract NAS5-26555.
MERLIN is a national facility operated by the University of Manchester on behalf of PPARC in the UK.
This research has made use of the NASA/IPAC Extragalactic Database (NED) which is operated by the Jet Propulsion Laboratory, California Institute of Technology, under contract with the National Aeronautics and Space Administration. 
We would like to thank Alison Peck and Greg Taylor for their contributions to this project. We also thank Paul Rayner for help with the MERLIN 5 GHz imaging, and acknowledge Dean Hines for his {\sc polarize} software. R. J. B. would like to acknowledge financial support from the European Commission's I3 Programme ``RADIONET'' under contract 505818. 
We thank the anonymous referee for useful and relevant comments and suggestions.

\newpage
\bibliographystyle{astron}
\bibliography{floyd_etal_3c293}

\begin{deluxetable}{lllrrr}
  \tabletypesize{\small}
  \tablecolumns{6}
  \tablewidth{0pc}
  \tablecaption{\label{tab-obs} Journal of Observations}
  \tablehead{\colhead{Instrument} & \colhead{Filter/Wavelength} &
    \colhead{Observing Date} & \colhead{$t$/s} &
    \colhead{Proposal ID} & \colhead{PI}}
  \startdata
  HST/NICMOS2 & {\sc F160W}  & $1998-08-19$ & 1695.8 & 7853 & Jackson\\
  HST/NICMOS2 & {\sc F170M}  & $1998-08-19$ & 3327.7 & 7853 & Jackson\\
  HST/NICMOS2 & {\sc F181M}  & $1998-08-19$ & 3327.7 & 7853 & Jackson\\
  HST/NICMOS2 & {\sc POL0L}  & $1998-08-19$ & 1663.9 & 7853 & Jackson\\
  HST/NICMOS2 & {\sc POL120L}& $1998-08-19$ & 1663.9 & 7853 & Jackson\\
  HST/NICMOS2 & {\sc POL240L}& $1998-08-19$ & 1663.9 & 7853 & Jackson\\
  MERLIN      & 6cm          & $1996-11-30$ & ...    & ...  & Leahy\\
  \multicolumn{6}{c}{\bf Archived observations}\\
  HST/STIS   & {\sc F25SRF2}& $2000-06-14$ & 1440.0 & 8275 & Sparks (1)\\
  HST/WFPC2  & {\sc F702W}  & $1995-01-15$ &  280.0 & 5476 & Sparks (2)\\
  MERLIN     & 18cm         & $1993-11-08$ &  ...   & ...  & (3)\\
  MERLIN     & 22cm         & $1998-04-08$ &  ...   & ...  & (4)\\
  \enddata
  \tablecomments{Details of the observations used in this paper. HST filter names are given for the optical-IR observations, wavelengths for the radio bands.}
  \tablerefs{(1)~\citet{allen+02}; (2)~\citet{martel+99}; (3)~\citet{akujor+96}; (4)~\citet{beswick+04}.}
\end{deluxetable}

\begin{deluxetable}{crrrrrr}
  \tabletypesize{\small}
  \tablecolumns{7}
  \tablewidth{0pc}
  \tablecaption{\label{tab-host-one} Single-component optical-infrared 
    host galaxy models}
  \tablehead{
    \colhead{Type} & \colhead{$\chi^{2}$} & \colhead{Flux/STMAG} & 
    \colhead{$R_{e}$/pix} & \colhead{$n$} & \colhead{$b/a$} & 
    \colhead{$\theta$}}
  \startdata
  Sersic & 2.01 & 14.00 & 447.0 & 3.93 & 0.57 & 67\\
  De Vauc& 2.05 & 12.50 &1589.9 & 4.00 & 0.56 & 68\\
  \multicolumn{7}{c}{\bf 1.6$\mu$m (F160W)}\\
  Sersic & 1.93 & 12.27 & 344.7 & 3.57 & 0.62 & 63\\
  De Vauc& 1.95 & 12.05 & 468.0 & 4.00 & 0.62 & 63\\
  \multicolumn{7}{c}{\bf 2.0$\mu$m (POL0L/POL120L/POL240L - total intensity)}\\
  Sersic & 1.82 & 12.24 & 165.0 & 3.32 & 0.62 & 60\\
  De Vauc& 1.90 & 11.97 & 246.6 & 4.00 & 0.62 & 60\\
  \enddata
  \tablecomments{Best-fit single-component host galaxy models at $R$, $H$ and $K$ band. Fluxes are given in STMAGs. Scale length, $R_e$ is measured in pixels on the final processed images (0\arcsec.045 for WFPC2; 0\arcsec.038 for NICMOS). The dimensionless S\'{e}rsic parameter, $n$ controls the shape of the profile, with $n=4$ corresponding to a de Vaucouleurs bulge, and $n=1$ to an exponential disk.}
\end{deluxetable}

\begin{deluxetable}{crrrrrr}
  \tabletypesize{\small}
  \tablecolumns{7}
  \tablewidth{0pc}
  \tablecaption{\label{tab-host-multi} Host Galaxy Models: $H$-band}
  \tablehead{
    \colhead{Type} & \colhead{Centre} & \colhead{$H$/STMAG} & 
    \colhead{$R_{e}/$pix} & \colhead{$n$} & \colhead{$b/a$} & 
    \colhead{$\theta$}}
  \startdata
  sersic & $(198.78,1477.13)$ & 17.49 & 37256.0 & 7.57 & 0.07 & 20\\
  sersic & $(559.14, 549.25)$ & 22.60 &   107.1 & 1.83 & 0.70 &-15\\
  sersic & $(553.28, 559.90)$ & 21.75 &   117.9 & 3.00 & 0.56 & 23\\
  \enddata
  \tablecomments{Multi-component host galaxy models ($H$-band). See table 2 for notes.}
\end{deluxetable}

\begin{deluxetable}{crrrrrr}
  \tabletypesize{\small}
  \tablecolumns{7}
  \tablewidth{0pc}
  \tablecaption{\label{tab-host-multicpt} Host Galaxy Models: $K$-band}
  \tablehead{
    \colhead{Type} & \colhead{Centre} & \colhead{mag.} & 
    \colhead{$R_{e}/$pix} & \colhead{$n$} & \colhead{$b/a$} & 
    \colhead{$\theta$}}
  \startdata
  sersic & $(388.10,1021.42)$ & 16.64 & 32300.1 & 3.04 & 0.11 & 16\\
  sersic & $(558.27, 553.88)$ & 23.87 &    65.9 & 0.96 & 0.71 &-12\\
  sersic & $(553.51, 559.57)$ & 22.07 &   172.6 & 4.17 & 0.56 & 21\\
  \enddata
  \tablecomments{Multi-component host galaxy model ($K$-band)}
\end{deluxetable}

\begin{deluxetable}{crrrrrrrrll}
  \tabletypesize{\small}
  \tablecolumns{10}
  \tablewidth{0pc}
  \tablecaption{\label{tab-SED}Jet SED}
  \tablehead{
    \colhead{Knot} & \colhead{$S(22~\mathrm{cm})$} & 
    \colhead{$S(18~\mathrm{cm})$} & \colhead{$S(6~\mathrm{cm})$} &
    \colhead{$S(2~\mu\mathrm{m})$} & \colhead{$S(1.6~\mu\mathrm{m})$} & 
    \colhead{$S(6940~\mathrm{\AA})$} & \colhead{$S(2299~\mathrm{\AA})$} &
    \colhead{$\alpha_{in}$} &  \colhead{$\nu_{b}$}\\

    \colhead{}&\colhead{mJy}&\colhead{mJy}&\colhead{mJy}&
    \colhead{$\mu$Jy}&\colhead{$\mu$Jy}&\colhead{$\mu$Jy}&\colhead{nJy}&
    \colhead{}&\colhead{$\times10^{13}Hz$}}
  \startdata
  E3 & 302.6 & 237.9 & 146.3 & 15.3 & 3.3 & 0.1 & 3.1 & $-0.61$ & 11.9\\ 
  E2 & 646.7 & 585.2 & 362.3 & 38.7 & 1.7 & ... & ... & $-0.50$ & 4.24\\
  E1 & 533.0 & 482.8 & 247.0 &  6.9 & 0.7 & ... & ... & $-0.65$ & 3.43\\
  \enddata
  \tablecomments{Knot flux in each band. Injection index and break frequency are given for the best-fitting \citet{jaffeperola73} radiation models, overlayed in Fig.~\ref{fig-SED}.}
\end{deluxetable}

\end{document}